\documentstyle[aps]{revtex}

\begin{document}
\title{SOLIDS WITH WEAK AND STRONG ELECTRON CORRELATIONS}

\author{P. Fulde}

\address{
Max-Planck-Institut f\"{u}r Physik komplexer Systeme,\\
N\"othnitzer Str.~38,
D-01187 Dresden (Germany)}

\date{\today}
\maketitle

\begin{abstract}

A number of methods are discussed which may serve for a treatment of
electron correlations in solids. When the electron correlations are
relatively weak like in semiconductors or a number of ionic crystals
one may start from a self-consistent field calculation and include
correlations by quantum chemical methods. An incremental computational
scheme enables us to obtain results of high quality for the ground
state of those systems. A number of examples demonstrates that
explicitely. 

Solids with strongly correlated electrons require the use of model
Hamiltonians. With their help one can tackle the problem of
determining spectral densities for those systems. The projection
technique is a useful tool here. In strongly correlated $f$ electron
systems electron or holes can crystallize with quite different
physical consequences as in the case of a Wigner crystal or Verwey
transition. Finally, different routes to heavy-fermion behavior are
discussed, another hallmark of strongly correlated electrons.

\end{abstract}

\newpage

\newcommand{\gsim}{\mathrel{\raise.3ex\hbox{$>$\kern-.75em\lower1ex\hbox{$\sim$\
}}}}
\newcommand{\lsim}{\mathrel{\raise.3ex\hbox{$<$\kern-.75em\lower1ex\hbox{$\sim$\
}}}}

\setcounter{equation}{0}
\section{Introduction}

The problem of electron correlations is one of the most fascinating
and challenging of condensed matter physics. It is also of great
practical importance. An example is the energy gap of a semiconductor
which we certainly would like to be able to calculate without any
adjustable parameter. Its size depends crucially on electron
correlations. Therefore we can hope to calculate it only if we are
able to treat correlations without making uncontrolled
approximations. Most people are unaware that this goal has not been
reached yet. For a detailed understanding and accurate treatment of
correlations it is mandatory to calculate the wavefunction of the
ground state and the excited states of the electronic system. There
are other powerful methods used in electronic theory which base their
advantages by avoiding just that, like the density functional theory
(DFT) and the local density approximation (LDA) to it [1-4]. The
latter has been extremely successful and useful from various points of
view. Its conceptional simplicity is particularly noteworthy. However,
by construction it excludes detailed insight into the correlation
problem which can be gained, e.g., by studying the pair-distribution
function $g(\underline{r}, \underline{r'})$ of the system. The latter
is defined by 

\begin{equation}
g(\underline{r}, \underline{r}') =
\frac{1}{\rho(\underline{r})\rho(\underline{r}')}
<\Phi\mid~\sum_{i\neq
j}~\delta(\underline{r'}-\underline{r}_i)\delta(\underline{r}-\underline{r}_j)
\mid\Phi>.
\end{equation}

Here $\mid\Phi>$ is the ground-state wavefunction of the system,
$\underline{r}_i,\underline{r}_j$ are the coordinates of the electrons
and $\rho(\underline{r})$ is their density at point $\underline{r}$.
This function yields the probability of finding an electron at point
$\underline{r'}$ provided there is one at point $\underline{r}$,
$relative$ to the one without that constraint. It is therefore
suitable for describing electrons which avoid each other in order to
reduce their mutual Coulomb repulsion. This gives rise to a
correlation hole which is associated with each electron. An accurate
description of this correlation hole, in particular at small distances
is the heart of the correlation problem. It is for that reason that
the pair-distribution function is intimately connected with
the exchange-correlation energy $E_{xc}[\rho]$ in density-functional theory.

\noindent The relationship is

\begin{equation}
E_{xc}~[\rho]=\frac{e^2}{2}\int~d^3r
d^3r'\rho(\underline{r})~\frac{\tilde{g}(\underline{r},\underline{r}')-1}{|\underline{r}-\underline{r}'|}~\rho(\underline{r}')
\end{equation}

where $\tilde{g}(\underline{r},\underline{r}')$ is related to
$g(\underline{r},\underline{r}')$ via

\begin{equation}
\tilde{g}(\underline{r},\underline{r}') = \int\limits^1_0
d\lambda~g(\underline{r},\underline{r}',\lambda). 
\end{equation}

\noindent The parameter $\lambda$ multiplies the interaction constant
$e^2$ of the electrons, which enters
$g(\underline{r},\underline{r}')$, i.e., $e^2\to \lambda e^2$ [5,6]
when $g(\underline{r},\underline{r}')$ is calculated. The simplicity
of the LDA or comparable approximations to density functional theory
results from making a reasonable $ansatz$ for
$g(\underline{r},\underline{r}')$ instead of trying to calculate it
from first principles. 						 

Given that one aims at calculating the many-electron wavefunction of a
periodic solid one is faced with a number of quantum-chemical methods
which are designed to do just that for small molecules. Particularly
well known are the Configuration-Interaction (CI) method [7], the
Coupled Electron Pair Approximation (CEPA) [8-10], and the Coupled
Cluster (CC) method [8,9,11,12]. These methods have in common that
they start out from a self-consistent field (SCF) Hamiltonian
$H_{SCF}$ and its ground-state wavefunction. Correlations are included
by admixing $SCF$ excited states to the ground state. For solids it is
important to use methods which are size consistent, i.e., which yield
a correlation energy proportional to the system size when the systems
are large. This excludes the CI method which lacks that property. It
is a variational method and therefore has the advantage of giving
upper bounds to the energy. The size-consistency problem can be
eliminated by expressing the energy in terms of cumulants
[13,14]. This ensures that all approximations remain size consistent. 

A SCF calculation will be a good starting point for a correlation
calculation only when the correlations are not too strong. This will
generally be the case for solids which involve $s$ and $p$ electrons
only. When $f$ electrons come into play, or in many cases also $d$
electrons, then the electron correlations are usually so strong that a
SCF wavefunction is not a good starting point for a computation of the
exact wavefunction. In a molecular calculation the problem may be
circumvented by starting instead from a multiconfiguration
self-consistent field calculation (MC-SCF). By a proper choice of the
active space (or of the different Slater determinants which are used)
one can include the most important correlations already at that level
and treat the remaining weaker ones by a subsequent CI or CEPA
calculation. For a solid this is not possible because several
configurations $per~lattice~site$ would be required, i.e., their
number would become infinitely large. Possibly, ways can be found in
the future to solve this problem. At present, however, we depend on
treating strongly correlated electrons by simplified model
Hamiltonians. Therefore we will subdivide here the discussion of
correlations into one of weakly and one of strongly correlated systems.

As regards weakly correlated systems another topic requires attention.
When applying methods like CI or CEPA in molecular calculations, the
excited states which are admixed with the SCF ground-state
wavefunction are usually the SCF ones, which extend generally over the
whole molecule. Going over to an infinite solid this would imply the
admixture of infinitely many excitations (configurations) to the
ground-state wavefunction $\mid\Phi_{SCF}>$. On the other hand, it is
evident that the correlation hole of an electron is a rather local
object, which is confined to the immediate surroundings of
it. Therefore, it should be described by $local$ operators acting on
$\mid\Phi_{SCF}>$ or $local$ excitations. This was first consequently
done in Ref. [15]. The same ideas were later applied in slightly
modified versions, e.g., in Refs. [16,17]. Working with local
operators is also useful in order to limit efficiently the number of
configurations one is admixing to $\mid\Phi_{SCF}>$. This brings us to
the concept of partitioning and projection which has been stressed by
L\"owdin [18]. Instead of constructing the $exact$ ground state
$\mid\psi_0>$ from $\mid\Phi_{SCF}>$ by applying operators chosen from
the full operator- or Liouville space $\Re$ one limits their choice to
a relevant subspace $\Re_0$. The latter must contain, e.g., all those
local operators which are necessary in order to describe the
correlation hole of the electrons with a required degree of
accuracy. Thus, instead of expanding the ground-state energy in powers
of the residual interactions $H_{res}= H-H_{SCF}$ one partitions $\Re$
into a relevant part $\Re_0$ and an irrelevant one $\Re_1$ and
projects the wave operators $\tilde{\Omega}$ defined by

\begin{equation}
\mid\psi_0>~=~\tilde{\Omega}\mid\Phi_{SCF}>
\end{equation}

\noindent onto $\Re_0$. Within that restricted space the ground-state
wavefunction and energy are found by diagonalization of the Hamiltonian.

The concept of projection and partitioning can also be applied to the
computation of Green's functions [19]. This is very useful when we
calculate the spectral density of a strongly correlated electron
system described by a model Hamiltonian. The ones of $Ni$ metal and of
$Cu-O$ planes, the most important structural element of the high $T_c$
superconducting cuprates serve as examples here. 

A particularly interesting phenomenon is that of electron or hole
crystallization. The conditions for its occurrance were first
determined by Wigner [20], who considered a homogeneous electron
gas. A prerequisite for electron crystallization is that the
electronic Coulomb repulsion dominates the kinetic energy, a condition
fulfilled for a homogeneous electron gas only when the density is very
low. In intermetallic rare-earth compounds the $4f$ electrons have a
small kinetic energy even at large densities because of the small
spatial extent of the $4f$ orbitals. Therefore, electron or hole
crystallization is expected to occur in favourable cases even at high
densities. There is clear evidence that $Yb_4As_3$ is an example of
$4f$ hole crystallization at low temperatures [21,22].

One much investigated feature of strongly correlated electron systems
is a high density of low-lying excitations. They involve predominantly
spin degrees of freedom and lead to heavy-fermion behaviour (for
reviews see [23-27]). There is a one to one correspondence between the
low-energy excitations of a heavy fermion system and those of a nearly
free electron system, provided one renormalizes the effective mass of
the latter. The characteristic low-energy scale of a heavy-fermion
system can be of different physical origin like the Kondo effect [28],
the Zeeman effect [29] or quasi one-dimensional spin chains [30]. The
materials $CeAl_3$ [31], $Nd_{2-x}Ce_xCu O_4$ [32] and $Yb_4 As_3$
[33] are examples for these three different physical cases. 

\setcounter{equation}{0} 
\section{Weakly correlated systems}

We start out describing the formalism used for the calculation of the
ground state and some of its properties of a weakly correlated
electron system. This is followed up by a presentation of the results
for a number of semiconductors and ionic crystals.  

\subsection{Projection techniques}

In order to ensure size consistency in the sense described above we
express all size-extensive quantities in terms of cumulants [13,14].

We divide the Hamiltonian $H$ into two parts

\begin{equation}
H ~ = ~ H_0+H_1
\end{equation}

\noindent and assume that the ground state $\mid\Phi_0>$ of $H_0$ is
known while $H_1$ must be treated approximately. In the case of
weakly-correlated electron systems we shall identify $H_0$ with the
self-consistent field Hamiltonian $H_{SCF}$ and $H_1$ with the
residual interactions, but we prefer keeping the discussion of the
projection technique more general. We write for the ground-state
energy

\begin{equation}
E_0~=~(H\mid\Omega)
\end{equation}

where

\begin{equation}
\Omega~=~\lim_{z\to 0} \left(1+\frac{1}{z-L_0-H_1}~H_1\right).
\end{equation}

Here the superoperator or Liouvillean $L_0$ refers to $H_0$, i.e., it
acts on operators $A$ according to

\begin{equation}
L_0 A~=~[H_0, A]_-.
\end{equation}

The brackets in Eq. (2.2) are defined as follows

\begin{equation}
(A\mid B)~=~<\Phi_0\mid A^+B\mid\Phi_0>^c.
\end{equation}

The upper script $c$ indicates that the cumulant of that expectation
value must be taken. The cumulant of a product of operators is defined by

\begin{equation} 
\left.
<A_1...A_n>^c~=~\frac{\partial}{\partial\lambda_1}...\frac{\partial}{\partial\lambda_n}\ell
n <\prod^n_{i=1}~e^{\lambda_i A_i}>\right|_{\lambda_1
=...=\lambda_n=0}. 
\end{equation}

For two operators this expression reduces to

\begin{equation}
<A_1 A_2>^c~=~<A_1A_2>-<A_1><A_2>.
\end{equation}

Using cumulants requires care [34]. For example, we must distinguish
between the number 1 and the unit operator $1_{op}$. The difference is

\begin{equation}
<1\cdot A>^c~=~<A>~~~{\rm and} ~~~<1_{op} A>^c~=~0.
\end{equation}

>From the properties of cumulants it follows that for any operator $A$
(but not for $c$ numbers) the following relation holds

\begin{equation}
(A|H\Omega)~=~0.
\end{equation}

\noindent The projection or partitioning method approximates
$|\Omega)$ by admitting for its construction only operators from a
reduced $relevant$ subspace $\Re_0$ of the full operator space
$\Re$. The operator $\Omega$ is thus projected onto $\Re_0$. The
quality of the approximation can be improved by successively
increasing the dimension of $\Re_0$. 

Let $\Re_0$ be of dimension $r_0$ and spanned by a set of operators
$\{A_\nu\}$ with $\nu=1,...,r_0$. We may choose the $A_\nu$ to be
orthonormal, i.e., 

\begin{equation}
(A_\nu|A_{\mu})~=~\delta_{\mu\nu}~.
\end{equation}

The wave operator is then given by

\begin{equation}
|\Omega)~=~|~1+\sum_{\nu}\eta_{\nu}A_{\nu}).
\end{equation}

In case that all $A_{\nu}$ couple directly to $H_1$, i.e., when

\begin{equation}
(A_\nu|H_1)\neq 0~~~{\rm for~all}~\nu
\end{equation}

one may use Eq. (2.9) to determine the coefficients $\eta_{\nu}$, i.e.,

\begin{eqnarray}
(A_{\mu}|H\Omega) & = & (A_{\mu}|H_1)+\sum_{\nu}\eta_{\nu}(A_{\mu}|H A_{\nu})\\ \nonumber
~                 & = & 0.
\end{eqnarray}

The solution of this set of equations can be written in the form

\begin{equation}
\eta_{\nu}~=~-\sum_{\mu}~L^{-1}_{\nu\mu}~(A_\mu|H_1)
\end{equation}

where

\begin{equation}
L_{\rho\tau}~=~(A_{\rho}|H A_{\tau}).
\end{equation}

The shift in the ground-state energy due to $H_1$ is given by

\begin{eqnarray}
\delta E_0 (\Re_0) & = & \sum_{\nu} \eta_{\nu}~(H_1|A_{\nu})\\ \nonumber
~                  & =  & \sum_{\nu}~\epsilon(\nu).
\end{eqnarray}

The energy shift consists of a sum of contributions $\epsilon(\nu)$
from the operators $A_\nu$ which span the reduced Liouville space
$\Re_0$. If $\{A_{\nu}\}$ contains operators which do not couple
directly to $H_1$ implying that $(A_\rho|H_1)=0$ for some $A_{\rho}$,
the latter enter only indirectly the calculations by modifying the
coefficients $\eta_{\nu}$. In that case they have the form of a
generalized continued fraction [13]. 

In order to give an example, we identify $H_0$ with the
self-consistent field Hamiltonian $H_{SCF}$ of $H$ and $H_1$ with the
residual interaction part $H_{res}$. We choose for the set
$\{A_{\nu}\}$ the single (S) and double (D) excitations out of the SCF
ground state $|\Phi_{SCF}>$. Single excitations are generated by
$\omega^i_{\mu}~=~c^+_i c_{\mu}$ where the $c_{\mu}$, $c^+_i$
operators destroy and create electrons in the molecular spin orbitals
$\phi_{\mu}$ and $\phi_i$, respectively. It is common to use Greek
indices for orbitals occupied in $|\Phi_{SCF}>$ and Latin indices for
unoccupied or virtual orbitals. Similarly, $\omega^{ij}_{\mu\nu}$
describes double excitations. By using compound indices $K$ and
$\Gamma$ we can write 

\begin{equation}
A^K_{\Gamma}\;=\; \left\{%
\begin{array}{l l l}
\omega^i_{\mu}& & \\[6pt] 
\omega^{ij}_{\mu\nu}&; & i<j~\quad \mbox{and}~\quad \mu<\nu.
\end{array}
\right. 
\end{equation}

The energy shift (2.16) is here equal to the correlation energy
$E_{corr}$ and the $\epsilon(\nu)$ are

\begin{equation}
\epsilon(\Gamma)~=~\sum_K~\eta^K_{\Gamma}(H_{res}|A^K_{\Gamma}).
\end{equation}

\noindent The above choice of the operator set $\{A_{\nu}\}$ and the
form (2.18) of the individual contributions to the correlation energy
are identical to the coupled-electron-pair-approximation ($CEPA-O$) in
quantum chemistry. 

It poses no problem to establish also a link between Eqs. (2.2, 2.9)
and the Coupled-Cluster method which was originally developed in
nuclear physics [12] but has been successfully applied also to quantum
chemistry [8,11]. As has been shown in Refs. [34,35] $|\Omega)$ can be
brought into the form 

\begin{equation}
\mid\Omega)~=~|e^S)
\end{equation}

where $S$ is a $prime$ operator, i.e, one which is treated as an entity when
cumulants are calculated. We may decompose $S$ into

\begin{equation}
S~=~\sum_\mu \eta_{\mu}S_{\mu},
\end{equation}

i.e., into a basis $S_{\mu}$ of prime operators which have the
additional property that $S_{\mu}|\Phi_0>\neq 0$. Prime operators $A$
with $A|\Phi_0> = 0$ need no further consideration. The $\eta_{\mu}$
are determined from $(S_{\nu}|He^S)=0$. Identifying the $S_{\mu}$ with
the $A^K_{\Gamma}$ (see Eq. (2.17)) we end up with the Coupled Cluster
approach with single and double substitutions ($CCSD$). 

\subsection{Incremental method}

We can perform correlation calculations of high accuracy for the
ground state of semiconductors or insulators applying the method of
increments [36-40]. 

Let us first consider the elemental semiconductors or the III-V and
II-VI compounds which all have well defined bonds. We express
$|\Phi_{SCF}>$ in terms of (localized) Wannier orbitals rather than
(delocalized) Bloch states. The Wannier orbitals define the different
bonds I. The relevant operators are identified with the operators
$A_I(n)$ and $A_{IJ}(m)$. The $A_I(n)$ describe one-and two-particle
excitations out of bond I where $n$ is a \-tcounting index. Similarly,
the $A_{IJ}(m)$ describe those two-particle excitations, where the
initial state of one excitation is the Wannier orbital I while that of
the other one is the Wannier orbital $J$. Therefore

\begin{equation}
|\Omega)~=~|1+\sum_{I,n}~\eta_I(n)A_I(n)+\sum_{<IJ>m}\eta_{IJ}(m)A_{IJ}(m))
\end{equation}

where the $\eta_I(n)$ and $\eta_{IJ}(m)$ are determined according to
Eq. (2.13). The method of increments provides a way of solving
approximately that system of equations. It is based on the following
idea. First, all electrons are kept frozen except those in bond
I. After the corresponding correlation energy $\delta
E_0~=~\epsilon(I)$ has been determined from Eq.(2.16), one repeats the
calculations by releasing electrons in bonds I and J. This yields a
new correlation energy $\delta E_0(I,J)$ or an increment 

\begin{equation}
\epsilon(I,J)~=~\delta E_0(I,J)-\epsilon(I)-\epsilon(J),
\end{equation}

as compared with the previous case. Thereafter, electrons in bonds
$I,J,K$ are released with the remaining ones kept frozen and so
on. After each step a new correlation energy $\delta E_0(I,J,K)$
etc. is obtained with corresponding increments  

\begin{eqnarray}
\epsilon(I,J,K) & = & \delta E_0(I,J,K) - \epsilon(I,J) - \epsilon(I,K)\\ \nonumber
~               & - & \epsilon(J,K) -\epsilon(I) -\epsilon(J) -\epsilon(K)
\end{eqnarray}

and so on. The total correlation energy $E_{corr}$ is then written in
terms of the energy increments as

\begin{equation}
E_{corr}=\sum_I\epsilon(I)~+\sum_{<IJ>}~\epsilon(I,J)+\sum_{<IJK>}~\epsilon(I,J,K)+...
\end{equation}

where $<IJ>, <IJK>$ denotes pairs and triples of bonds,
respectively. The advantage of this expansion is that the series is
usually rapidly convergent [36] even when $H_{res}$ is not small. It
has been demonstrated that there are strong relations between the
method of increments and Faddeev's equations, provided the latter are
properly generalized to a hierarchy of cumulant equations [34]. These
relations enable us to state the size of the error when the series
(2.24) is terminated at some point. 

When ionic crystals are considered the increments refer to different
ions instead of bonds. For example, the first increment to be
calculated for $Mg O$ is the correlation energy attributed to an
$O^{2-}$ ion surrounded by $Mg^{2+}$ and $O^{2-}$ ions. An accurate
modelling of the surroundings is essential, since an isolated $O^{2-}$
is not stable. Similarly, the correlation-energy increment $Mg\to
Mg^{2+}$ has to be determined. The two-body increments involve pairs
of $O-O$, $O-Mg$ and $Mg-Mg$ ions and so on. 

An important point is the following. The matrix element
$(H_{res}|A^K_{\Gamma})$ and $(A^{K'}_{\Gamma'}|H A^K_{\Gamma})$
which are needed to determine the correlation energy (see Eq. (2.18))
can be computed from clusters. When the cluster is chosen so large
that a matrix element does not change any more, we are sure that its
value is that for the solid. This enables us to apply quantum chemical
program packages like MOLPRO [41] to determine them and hence the
different correlation contributions.

\subsection{Results for semiconductors and ionic crystals}

Starting point for the computation of the ground-state wavefunction
and energy is a $SCF$ calculation. Obtaining accurate $SCF$ energies
for solids is still a difficult computational task. Pisani and
co-workers have developed the successful program package CRYSTAL [42]
which allows for $SCF-LCAO$ (linear combination of atomic orbital)
calculations on periodic solids. It expands the Bloch orbitals in
terms of Gaussian-type orbitals (GTO's) and obtains the canonical SCF
solutions. The latter can always be transformed to a localized
representation by using a suitable localization procedure
[43]. However, convergency problems are encountered when Gaussian
functions with small exponents are used. An alternative ab-initio
embedded-cluster approach has been developed within the framework of
the LCAO method in which the localization procedure is part of the
process leading to the SCF solutions [44]. Thereby the solid is
modelled as a central cluster embedded in the field created by the
remaining part of the infinite solid. Test calculations for $LiH$
agree with those obtained from the CRYSTAL program, but large scale
applications have not been performed yet.

A third possibility for obtaining the SCF ground-state wavefunction of
a solid is by means of cluster calculations. A fragment of the lattice
is used with dangling bonds saturated by hydrogen atoms. An example is
$X_{26}H_{30}(X=C, Si,Ge, Sn)$. The calculated localized orbital for
the central bond of the cluster can be used to very good approximation
for every bond of the solid. 

After the SCF ground-state wavefunction has been determined the
correlation corrections are incorporated either on a $CEPA-O$ (see
Eqs. (2.17-18)) or $CCSD$ (see Eqs. (2.19-20)) level, thereby applying
the method of increments. We discuss first the results for the
elemental semiconductors and the cubic III-V compounds before we turn
our attention to the ionic crystals.\\ 

\noindent\underline{\bf Semiconductors}

The $SCF$ calculations for the valence electrons are done with a basis
set consisting of $GTO's$. Its size is $(4s 4p 1d)/[3s 3p 1d]$ in standard
notation, i.e., four $s$-like GTO's are contracted into three orbitals
and similarly for $p$-like functions. The inner shells are described
by a quasirelativistic pseudopotential [45]. In some cases a
polarization potential is added in order to account for the
polarization of the inner shells by the valence electrons. For further
details concerning the basis set we refer to the original literature
[46]. The cohesive energy obtained within the SCF approximation is
only between 55-72 \% for the elemental semiconductors and between 55
- 67 \% for the cubic III-V compounds (see Tables 1,2). Thus, even for
these weakly correlated systems the correlation contribution to
binding is appreciable. Lattice constants come out much better in SCF
approximation, the errors being less than 2\% (see Fig. 3). Bulk
moduli\-li are generally overestimated by up to 20 \% (see, e.g., Table
4). For a treatment of correlations the basis sets have to be
enlarged. Instead of using one set of $d$ functions only, we use two
sets of $d$ functions and one set of $f$ functions. The basis is
therefore of size $(4s 4p 2d 1f)/[3s 3p 2d 1f]$ which is very respectable for
diamond and silicon, but less for heavy elements like $Ge$ or $Sn$
because of low lying $d$ states. Results for the cohesive energy of
the group IV semiconducting and the cubic III-V compounds are listed
in Tables 1 and 2, respectively. One notices that about 90 \% of the
binding energy are obtained in all cases. The missing percentage is
due to basis set limitations. By including a second set of $f$
functions and one set of $g$ functions one would further improve the
accuracy of the calculated values, in particular for the systems
involving heavy atoms. 

It is well known that bond lengths or alternatively lattice constants
are relatively well reproduced by a SCF calculation. The same holds
true for the group IV semiconductors as is seen from Table 3. The
inclusion of a core polarization potential is important in order to
come much closer to the experimental values. When correlations among
valence electrons are included in the way described above the error
for the lattice constant is less than 0.7 \% in all cases. 

The bulk modulus $B$ is defined by

\begin{equation}
B = \left( \frac{4}{9 a}\frac{d^2}{d
a^2}~-~\frac{8}{9a^2}\frac{d}{da}\right)E_0(a) 
\end{equation}

where $E_0(a)$ is the ground-state energy as function of the lattice
parameter $a$. As mentioned before, self-consistent field calculations
overestimate generally $B$. The core-polarization potential as well as
the correlation energy lead to a decrease of $B$ so that the final
results are close to the experimental ones (see Table 4). This is easy
to understand: when correlations are taken into account they keep
electrons apart. Therefore an increase in pressure results in a
smaller energy increase than obtained by a SCF treatment, where
electrons come too close to each other at the expense of Coulomb
repulsion energy. 

We are particularly interested in trends concerning the $inter$atomic
correlation energy. The latter is obtained when a minimal basis set is
used, i.e., one containing one set of $s$ and $p$ functions
only. Correlations, which require for their description a larger than
minimal basis set are termed $intra$-atomic [14]. Interatomic
correlations reduce charge fluctuations between neighboring atoms. The
simplest way of obtaining a good estimate of the one-bond and two-bond
contributions is by making a bond-orbital approximation (BOA). In
this approximation the ground state is written as a product of
independent bonds containing two electrons each. In that case the
interatomic correlation energy depends on the following quantities:
the interaction matrix element $V_0$ describing the difference of the
Coulomb repulsion when two electrons are situated on the same atom and
when they are placed on different atoms of the same bond; the
electronic hopping matrix element $t_0$ between atoms of the same
bond; the van der Waals interaction, in simplest approximation between
neighboring bonds, $V_1$; the bond polarity $\alpha$ in case of
heteropolar bonds [52]. 

For the elemental semiconductors the bare hopping matrix element $t_0$
scales with the tight-binding band width of the system. For diamond
the value is $t^C_0=10.7 eV$ as obtained from band-structure
calculations. We can determine the one for $Si$, $Ge$ and $\alpha-Sn$
according to Froyen and Harrison [53] from the width of the LCAO
bands. Once $t_0$ is known the parameter $V_0$ and $V_1$ follow from
the one-bond and two-bonds increments  

\begin{eqnarray}
\epsilon(I)   & = & -\frac{V_0}{4t_0}(1-\frac{6V^2}{t_0 V_0})\\ \nonumber
\epsilon(I,J) & = & -\frac{V_1^2}{t_0}(1-\frac{V_0}{2t_0});
\end{eqnarray}

(I and J label nearest-neighbor bonds) when the latter are computed
with a minimal basis set. A list of the calculated parameters is given
in Table 5. One can use them in order to derive a simple scaling
relation between the size of $t_0,V_0,V_1$ and the bond length
$d$. One finds that the following relations describe the data fairly
well [54] 

\begin{eqnarray}
t^0/t^C_0 & = & (d_C/d)^{1.6}\\ \nonumber
V_0/V^C_0 & = & (d_C/d)^{1.7}\\ \nonumber
V_1/V_1^C & = & (d_C/d)^{1.4}
\end{eqnarray}

where the script $C$ refers to diamond. Going over to the polar III-IV
compounds one expects a relationship

\begin{equation}
E^{inter}_{corr}(\alpha)~=~(1-\alpha^2)^{\nu} E^{inter}_{corr}(0)
\end{equation}

\noindent between the interatomic correlation energy in the presence
of the polarity $E^{inter}_{corr}(\alpha)$ and the one calculated from
$t_0,V_0,V_1$ without taking the polarity into account, i.e.,
$E^{inter}_{corr}(0)$. In fact, the BOA predicts that $\nu=5/2$. But,
when one calculates $E^{inter}_{corr}(0)$ from Eqs. (2.28) using the
experimental values of $d$ and compares it with the $ab~initio$
results using minimal basis sets one finds a value of $\nu = 4.0 \pm
0.25$. The difference between the two approaches is the following. In
the BOA the excited electrons remain in their original bonds while in
the ab inito approach excitations into other bonds are taken into
account too. A simple estimate of the $intra$atomic correlation energy
is more difficult. Using the values of the free atoms is a too rough
method. One way is to apply an $atoms~in~molecules$ approach [55]
originally suggested by Moffitt. It takes into account that the ratio
of $s$ and $p$ electrons differs from that of a free atom when a solid
is formed. Also, the electron number fluctuates when the atom is part
of the solid and this is also accounted for in that approach. For more
details we refer to the literature [14].\\ 

\noindent\underline{\bf Ionic crystals}

The calculations for the ground state of semiconductors described above
can be extended to ionic crystals. Also here, correlations contribute
approximately 1/3 of the cohesive energy. We demonstrate this by
studying $Mg O, CaO$ and $NiO$. Thereby we employ again the method of
increments, but in distinction to the semiconductors considered
before, the increments refer here to ions instead of bonds. SCF
calculations for the binding energy, lattice constant and bulk modulus
were done by applying the program package CRYSTAL. Results are found
in Tables 6 and 7. 

Correlation calculations are performed on clusters of one, two and
three ions which are embedded in their proper surroundings. The
$X^{2+}~(X=Mg,Ca,Ni)$ ions are described by small-core
pseudopotentials and extended basis sets or a large-core
pseudopotential together with a small basis set and a
core-polarization potential. Test calculations for single ions show
that the potentials describe very well the results of $CCSD$
calculations with all-electron basis sets. The latter are in good
agreement with the experimental values for the first and second
ionization potentials $X\rightarrow X^+,X^+\rightarrow X^{2+}$, in
particular when they are supplemented by a perturbational treatment of
the triplet excitations (CCSD(T)). For $NiO$ we applied quasi
degenerate variational perturbation theory (see Ref. [60]) for
technical reasons: the MOLPRO program package does not allow for
$CCSD$ calculations of $NiO$ with low spin.

The one-body increments are attributed to an $O^{2-}$ and a $X^{2+}$
ion. Consider first the $O^{2-}$ ion. It must be embedded in a proper
surroundings, since a free $O^{2-}$ ion does not exist. Stabilization
is achieved  describing the six nearest neighbors by means of a
pseudopotential. The crystal environment of the resulting seven-atom
cluster is represented by 336 ions with point charges $\pm~2$
surrounding this cluster in form of a 7x7x7 cube. Employing a basis
set [$5s4p3d2f$] one obtains the increment $O\to O^{2-}$ shown in
Table 8. It is noticed that it is nearly independent of the system
considered. This is different for the increment $X\to X^{2-}$ which
increases from $Mg$ to $Ca$ to $Ni$ (see Table 8). This can be
understood by realizing that for $Ca$, excitations into the low lying
$d$ orbitals are important while for $Ni$ the correlation energy is
further enhanced by near degeneracies of $d$ orbitals. A more detailed
picture of the one-body increments for $NiO$ is shown in Fig. 1. For
$MgO$ and $Ca O$ the results are qualitatively similar. 

Of particular interest are two-body increments. Hereby we have to
distinguish between $O-O,~X-O$ and $X-X$ increments. When the distance
between the two sites becomes large, then in all three cases the
increments describe van der Waals interactions of the ions. We show in
Fig. 2 the $O-O$ increments for $Mg O$ for different distances. They
do not contain their respective weight factors. Table 9 lists the different
two-body increments for the three oxides.

Van der Waals interactions can be estimated with the help of London's
formula. It states that the lowering of the energy of two sites $A$
and $B$ due to van der Waals interactions is given by [62]

\begin{equation}
\Delta E \simeq -\frac{3}{2} \eta~\frac{I_A I_B}{I_A+I_B}~\frac{\alpha_A\alpha_B}{R^6}.
\end{equation}

where $I_{A(B)}$ are the ionization energies, $\alpha_{A(B)}$ are the
polarizability volumes, $R$ is the distance between sites $A$ and $B$
and $\eta$ is a constant of order unity. It relates a proper atomic
mean-excitation energy to the ionization energy. In Fig. 3 we have
plotted the calculated values for the van der Waals interaction
between positive and negative ions (circles) and between the
negatively charged ions (crosses) when quantum-chemical methods are
applied (here CCSD) and when London's formula is used. The data can be
roughly approximated by a straight line of slope 4. This suggests that
Eq. (2.29) should be multiplied by a correction factor of order 4.

Three-body increments are very small. For the three oxides the sum of
them is of order 0.03 eV~($MgO$) 0.05 eV~($Cu O$) and 0.11 eV ($Ni
O$). In Fig. 4 a summary of the different increments to the cohesive
energy of $MgO$ is given. Table 6 summarizes the results for the
cohesive energy for all three oxides. Table 7 does the same for the
calculated lattice constants. It is found that correlations have two
effects on the latter. The van der Waals interactions lead to a
reduction of the lattice constant since the corresponding energy gain
increases with decreasing distances. On the other hand, the
correlation energy of $O^{2-}$ increases with increasing lattice
spacing since the excitation energies are lowered in that
case. Therefore, an increase of the lattice constant is favoured by
that effect. This is opposite to what one would expect from the LDA
because the density decreases as the lattice constant increases. We
find that the calculated lattice constants deviate by less than 1 \%
from the experimental values. 

Ground-state calculations based on quantum chemical methods are
certainly more costly than those based on approximations to
density-functional theory. However, they allow for systematic
improvements by using larger basis sets. For the systems discussed
here the quality of the results is certainly comparable with that of
the generalized gradient corrected LDA.

In addition to the ground-state calculations one would like to be able
to calculate the energy band for a system like $Ni O$. For work in
this direction see, e.g., Refs. [64, 65] and further references cited
there. 

\setcounter{equation}{0}
\section{Strongly correlated electron systems}

In solids with $d$- or $f$- electrons the electron correlations are
generally strong. In that case a SCF calculation is not a good
starting point. This holds particularly true for $4f$ or $5f$ systems,
but also a treatment of the electrons in the $Cu-O$ planes of the
high-$T_c$ superconducting cuprates should include from the outset the
strong on-site Coulomb interactions of holes in the $Cu~3d$
shell. Calculations of this kind can be presently done only with
simplified model Hamiltonians. They all have in common that the basis
set is a minimal one, or even less, like in the case of a one-band
Hubbard model, which has been proposed for a study of $d$
electrons. From quantum-chemical calculations on small systems it is
known, that a minimal basis set gives unsatisfactory results in almost
all cases. Nevertheless, the hope has been that an understanding of
the model systems will yield generic properties of strongly correlated
electron systems which should prevail when eventually a more realistic
description should become possible one day. This applies, in
particular, to the low-lying excitations in those systems. It should
be pointed out that the local-density approximation to the
density-functional theory does not have basis set problems. Since the
contributions to the energy are calculated from an
exchange-correlation expression for a homogeneous electron gas,
characteristic features of strongly correlated electrons as obtained,
e.g., from a Hubbard Hamiltonian are not present in that
approach. This changes partially when a LDA+U scheme is used [66],
where an on-site repulsion term $U$ is added by hand to the systems
energy. First, we describe the Hamilton operators most commonly
used. Afterwards, we use them in order to calculate the spectral
density for $Ni$ with special emphasis put on the reproduction of the
observed satellite peak in photoelectron spectroscopy [67], and of
doped $Cu-O$ planes which are part of the cuprates. 

\subsection{Model Hamiltonians}

The most frequently used model Hamiltonian for strongly correlated
electron systems, is the one first suggested independently by
Gutzwiller [68], Hubbard [69] and Kanamori[70] and commonly referred
to as Hubbard Hamiltonian. It is of the form
 
\begin{equation}
H = - t \sum_{<ij>\sigma}~(a_{i\sigma}^+ a_{j\sigma}+h.c.)~+~U\sum_i~n_{i\uparrow}n_{i\downarrow}.
\end{equation}

The first term denotes the kinetic energy with $<ij>$ referring to a
pair of nearest-neighbor sites. The second term refers to the on-site
Coulomb repulsion with $n_{i\sigma}=a^+_{i\sigma}a_{i\sigma}$. For
strongly correlated systems the ratio $t/U\ll~1$. In that limit the
Hamiltonian can be transformed into one acting on a reduced Hilbert
space from which all configurations containing doubly occupied sites
$i$ are excluded. This effective Hamiltonian is of the form [71] 

\begin{equation}
H_{t-J}~=~-t\sum_{<ij>\sigma}~(\hat{a}_{i\sigma}^+\hat{a}_{j\sigma}+h.c.)+J\sum_{<ij>}
(\underline{S}_i\underline{S}_j-\hat{n}_i\hat{n}_j)
\end{equation}

\noindent and is called $t-J$ Hamiltonian. The
$\hat{a}^+_{i\sigma},\hat{a}_{i\sigma}$ operators act on the reduced
Hilbert space only and are given in terms of the
$a_{i\sigma}^+,a_{i\sigma}$ operators as

\begin{eqnarray}
\hat{a}^+_{i\sigma} & = & a^+_{i\sigma}(1-n_{i-\sigma})\\ \nonumber
\hat{a}_{i\sigma}   & = & a_{i\sigma}(1-n_{i-\sigma}).
\end{eqnarray}

The spin operators are
$\underline{S}_i=(1/2)\sum_{\alpha\beta}\hat{a}^+_{i\alpha}
\underline{\sigma}_{\alpha\beta}\hat{a}_{i\beta}$ where
$\underline{\sigma}$ refers to the Pauli matrices. Furthermore,
$\hat{n}_i=\sum_{\sigma}\hat{a}_{i\sigma}^+\hat{a}_{i\sigma}$ and the
coupling constant is $J=4t^2/U$. For transition metals a five-band
Hamiltonian describing the rather strongly correlated $d$ electron is
often used. It is of the form 
 
\begin{eqnarray}
H         & = & H_0 +\sum_{\ell} H_1(\ell)\\ \nonumber
H_0       & = & \sum_{\nu\sigma\underline{k}}~\epsilon_{\nu}(\underline{k})n_{\nu\sigma}(\underline{k})\\ \nonumber
H_1(\ell) & = & \frac{1}{2} \sum_{ijmn}\sum_{\sigma\sigma'} V_{ijmn}a^+_{i\sigma}(\ell) a^+_{m\sigma'}(\ell)a_{n\sigma'}(\ell)a_{j\sigma}(\ell)
\end{eqnarray}

where $\ell$ is a site index and $i,j,m,n$ label different $d$ orbitals.
The $\epsilon_{\nu}(\underline{k})~(\nu=1,...,5)$ describe the energy
dispersions of the canonical $d$ bands. Furthermore,
$n_{\nu\sigma}(\underline{k})=c^+_{\nu\sigma}(\underline{k})c_{\nu\sigma}(\underline{k})$
The creation operators $c^+_{\nu\sigma}(\underline{k})$ of the Bloch
eigenstates are expressed in terms of the $a^+_{i\sigma}(\ell)$ as

\begin{equation}
c^+_{\nu\sigma}(\underline{k})~=~\frac{1}{\sqrt{N}}\sum_{i\ell}~\alpha_i(\nu,
\underline{k})a^+_{i\sigma}(\ell)e^{i\underline{k}\underline{R}_{\ell}}
\end{equation}

\noindent ($N$ is the number of sites). The interaction matrix
elements are of the form 

\begin{eqnarray}
V_{ijmn} & = & U_{im}\delta_{ij}\delta_{m n}+J_{ij}(\delta_{in}\delta_{jm}+\delta_{im}\delta_{jn})\\ \nonumber 
U_{im}   & = & U + 2J - 2J_{im}
\end{eqnarray}

\noindent where $U$ and $J$ are average Coulomb- and
exchange-interaction constants. For a cubic system the matrix $J_{ij}$
can be expressed in terms of the average exchange constant and a
single anisotropy parameter $\Delta J$ only. The explicit form of the
matrix is found, e.g., in Ref. [72]. 

For a description of electrons in the $Cu-O$ planes, the crucial
structural element of the high-$T_c$ cuprates, a three-band Hubbard
model is usually used. It takes into account only the
$Cu~3d_{x^2-y^2}$ and the $O~2p_{x(y)}$ orbitals (see Fig. 5). Their
orbital energies are $\epsilon_d$ and $\epsilon_p$. Two holes on a
$Cu$ or $O$ site interact with an on-site Coulomb matrix element $U_d$
and $U_p$, respectively. The hopping matrix element between a $Cu$
orbital and an $O$ orbital is denoted by $t_{pd}$. Values for the
different parameters can be obtained from constrained LDA calculations
[73]. There is general consent that the following values are
reasonable: $U_d=10.5 eV, U_p=4.0 eV,~t_{pd} =1.3 eV, t_{pp}=0.65 eV,
\epsilon_p - \epsilon_d = 3.6 eV$ (in hole representation). 

Written in a basis of $O$ orbitals which is diagonal with respect to
$O-O$ hopping, the three-band Hubbard Hamiltonian reads

\begin{eqnarray}
H & = &
\sum_{m\underline{k}\sigma}~\epsilon_m(\underline{k})p^+_{m\underline{k}\sigma}p_{m\underline{k}\sigma}
+U_p\sum_J n_{p\uparrow}(J)n_{p\downarrow}(J)~+\\ \nonumber ~ & + &
\epsilon_d \sum_{\underline{k}\sigma}
d^+_{\underline{k}\sigma}d_{\underline{k}\sigma}+U_d\sum_I
n_{d\uparrow}(I)n_{d\downarrow}(I)~+\\ \nonumber ~ & + &
2t_{pd}\sum_{m\underline{k}\sigma}(\phi_{m\underline{k}}p^+_{m\underline{k}\sigma}d_{\underline{k}\sigma}+
\phi^*_{m\underline{k}}d^+_{\underline{k}\sigma}p_{m\underline{k}\sigma}).
\end{eqnarray}

\noindent Here $I$ and $J$ are indices for the $Cu$ and $O$ sites,
respectively. Furthermore, $n_{p\sigma}(J)=p^+_{J\sigma}p_{J\sigma}$
and $n_{d\sigma}(I)=d^+_{I\sigma}d_{I\sigma}$. The
$\epsilon_m(\underline{k})$ are given by 

\begin{eqnarray}
\epsilon_m(\underline{k}) & = & \epsilon_p\pm 2t_{pp}[cos \underline{k}(\underline{r}_1+\underline{r}_2)-cos
\underline{k}(\underline{r}_1 - \underline{r}_2)]\\ \nonumber
                     (  m & = & 1,2)
\end{eqnarray}

\noindent with vectors $\underline{r}_1, \underline{r}_2$ pointing
from a $Cu$ site to the two $O$ sites of the unit cell. The phase
factors $\phi_{m\underline{k}}$ are 

\begin{equation}
\phi_{m\underline{k}}~=~\frac{-i}{\sqrt{2}}~[{\rm sin}~\underline{k}\underline{r}_1~\pm {\rm sin}~\underline{k}\underline{r}_2]. 
\end{equation}

\noindent The model Hamiltonians (3.4) and (3.7) are used in the
following in order to calculate the spectral density of $Ni$ and of
the $Cu-O$ planes, respectively. In both cases the Coulomb repulsion
exceeds the hopping-matrix elements and therefore a simple
perturbation expansion in terms of the interactions is not
sufficient. Instead, the large interactions must be taken into account
more accurately. This goal is achieved by applying projection
techniques. 

\subsection{Spectral densities}

Spectral densities can be measured either by photoelectron emission
(PES) or inverse photoemission spectroscopy (IPES). They are
calculated from the positions and intensities of the poles of the
single-particle Green's function. One way of determining those poles
is treating numerically small clusters, e.g., by means of the Lanczos
method [74]. Thereby periodic boundary conditions are assumed. Here we
shall proceed differently. Using projection and partitioning
techniques we are able to do the calculations analytically to a large
extent with some numerical work remaining. The latter is by far less
than the one in the numerical approaches mentioned before.

We start out describing the projection technique as applied to the
calculation of spectral densities. Consider a set of wave-number
dependent operators $A_m(\underline{k})$ in Heisenberg representation,
i.e., with a time dependence given by 

\begin{equation}
A_m(\underline{k},t)=e^{it(H-\mu N)}A_m(\underline{k}) e^{-it(H-\mu N)}.
\end{equation}

\noindent Here $H$ denotes the Hamiltonian of the system while $\mu$
is the chemical potential and $N$ is the operator of the total
electron number. Using a grand-canonical ensemble and working with a
fixed value of $\mu$ has the advantage that the calculations can be
readily extended to finite temperatures. 

We define a matrix of retarded Green's functions
$G_{mn}(\underline{k},t)$ for the set of operators $A_m(\underline{k})$ through
  
\begin{equation}
G_{mn}(\underline{k},t)=-i\theta(t)<\psi_0|[A^+_m(\underline{k},t),A_n(\underline{k})]_+|\psi_0>
\end{equation}

\noindent where $\psi_0$ is the ground state of the system described
by $H$ while $\theta(t)$ is the step function which equals 1 for $t>0$
and $0$ for $t\leq 0$. Introducing the bilinear form 

\begin{equation}
(A|B)_+ = <\psi_0|[A^+,B|_+|\psi_0>
\end{equation}

\noindent we can write the Laplace transform of Eq. (3.11) in the form

\begin{equation}
G_{mn}(\underline{k},z)=(A_m(\underline{k})|\frac{1}{z-L}A_n(\underline{k}))_+.
\end{equation}

\noindent The Liouvillean $L$ acts on operators $A$ according to

\begin{equation}
LA~=~[H,A]_-.
\end{equation}

\noindent The spectral functions $A_{mn}(\underline{k},\omega)$
belonging to the set of operators $\{A_m(\underline{k})\}$ are defined
according to 

\begin{equation}
A_{mn}(\underline{k},\omega)~=~-\frac{1}{\pi}\lim_{\eta\to 0}Im\{G_{mn}(\underline{k},\omega+i\eta)\}.
\end{equation}

\noindent Equation (3.13) is in a proper form for applying projection and
partioning techniques. The idea of partitioning was put forward by
L\"owdin [18] and applied in particular by Pickup and Goscinski [75]
and Linderberg and \"Ohrn [76]. Projection techniques were also
introduced by Mori and Zwanzig for the description of dynamical
correlation functions. Here we will use an extension of the technique
to static quantities in a form which is size consistent [13,14].

We proceed as follows: We add (or remove) an electron in a Bloch state
of the system  for which we want to compute the spectral density. One
of the $A_m(\underline{k})$ is identified with the corresponding
creation (annihilation) operator. The strong correlations are taken
into account by a proper choice of local operators to which the added
electron (hole) couples strongly. They modify the surroundings of the
electron added to the probe (IPES) or of the hole left behind by an
ejected electron (PES). We identify the most important of these
processes from case to case and, after a Fourier transformation,
include the corresponding operators in the set
$\{A_m(\underline{k})\}$. Examples are given later when the theory is
applied to $Ni$ and the $Cu-O$ planes. All other microscopic processes
are discarded. The operator space $\Re$ is partitioned this way into a
$relevant$ part $\Re$ spanned by the $\{A_m(\underline{k})\}$ and an
orthogonal, $irrelevant$ part $\Re_1$ which is neglected. The desired
spectral density is then obtained from one of the diagonal elements of
the Green's function matrix $G_{mn}(\underline{k},\omega + i\eta)$
(see Eq. (3.15)). We rewrite Eq. (3.13) in matrix notation as

\begin{equation}
G(\underline{k},z)~=~\underline{\underline{\chi}}(\underline{k})[z\underline{\underline{\chi}}(\underline{k})- 
\underline{\underline{\omega}}(\underline{k})]^{-1}\underline{\underline{\chi}}(\underline{k})
\end{equation}

\noindent with the susceptibility and frequency matrix defined by

\begin{eqnarray}
\chi_{mn}(\underline{k})   & = & (A_m(\underline{k})|A_n(\underline{k}))_+\\ \nonumber
\omega_{mn}(\underline{k}) & = & (A_m(\underline{k})|LA_n(\underline{k}))_+~,
\end{eqnarray}

\noindent respectively. There is no memory matrix appearing in
Eq. (3.15) because we neglect $\Re_1$.

The matrix elements (3.17) can be evaluated in two different ways. One
consists in using a general relationship between spectral functions
and static expectation values. For the present case it can be written
in the form 

\begin{equation}
<\psi_0|A^+_m(\underline{k})A_n(\underline{k})|\psi_0>~=~\int\limits^{+\infty}_{-\infty}~d\omega A_{mn}(\underline{k},\omega)f(\omega).
\end{equation}

\noindent Here $f(\omega)$ is the Fermi function which can be replaced
by a step function. With the help of this relation the static
expectation values (3.17) can be determined self-consistently. The
second way of evaluating them is by applying the projection technique
also to these static quantities. Thereby one uses the following relations

\begin{eqnarray}
<\psi_0|[A^+_m(\underline{k}), A_n(\underline{k})]_+|\psi_0>  & = & (\Omega|[A^+_m(\underline{k}),A_n(\underline{k})]_+\Omega)\\ \nonumber
<\psi_0|[A^+_m(\underline{k}), LA_n(\underline{k})]_+|\psi_0> & = & (\Omega|[A^+_m(\underline{k}),(LA_n(\underline{k}))^{\cdot}]_+\Omega)
\end{eqnarray}

\noindent which are derived, e.g., in Ref. [14]. The round brackets
and $\Omega$ are defined by Eqs. (2.5) and (2.3), respectively. The
notation $(A B)^{\cdot}$ implies treating the product $AB$ as an
entity when the cumulant is calculated. This completes the description
of the projection technique as applied to Green's function
calculations. When the theory is applied, a proper choice of the
operator set $\{A_m(\underline{k})\}$ and of $\Omega$ is crucial.

\subsection{Application to $3d$ transition metals}

We apply the above theory in order to calculate the direct and inverse
photoemission spectra of $3d$ transition metals with special reference
to $Ni$. It has been known for a long time that electronic
correlations influence considerably the excitation spectra of these
materials. A well studied case is that of $Ni$, where angular-resolved
photoemission data reveal a 25 \% reduction in bandwidth as compared
with LDA bandstructure calculations and also the appearance of a
satellite structure 6 eV below the Fermi energy [77]. A number of
different investigations have dealt with these experimental
findings. Starting with the work of Kanamori [70], different methods
have been applied by Penn, Liebsch, Igarashi, Roth, Hertz and Edwards
and others [78-82]. We will show that the projection technique is here
a valuable tool leading to satisfactory results for $Ni$ [83].

We start from the Hamiltonian (3.4) for the $3d$ electron and identify
the $\epsilon_{\nu}(\underline{k})$ with the canonical $d$ bands
obtained, e.g., from a LDA calculation. The spectral density is
obtained from the retarded Green's functions

\begin{equation}
G_{\nu\sigma}(\underline{k},t)=-i\Theta(t)<\psi_0|[c_{\nu\sigma}(\underline{k},t),c^+_{\nu\sigma}(\underline{k},0)]_+|\psi_0>.
\end{equation}

We project again onto the relevant part $\Re_0$ of the Liouville space
spanned by a set of operators $\{A_m(\underline{k})\}$, one of which
is $c^+_{\nu\sigma}(\underline{k})$. The associated Green's function
matrix is again of the form of Eq. (3.13) which is rewritten as in
Eq. (3.16). The susceptibility and frequency matrix are of the form of
Eq. (3.19), i.e., 

\begin{eqnarray}
\chi_{mn}(\underline{k})   & = & (\Omega|[A^+_m(\underline{k}), A_n(\underline{k})]_+\Omega)\\ \nonumber
\omega_{mn}(\underline{k}) & = & (\Omega|[A^+_m(\underline{k}), (LA_m(\underline{k}))^{\cdot}]_+\Omega).
\end{eqnarray}

The ground state $\mid\Phi_0>$ onto which $\Omega$ is acting is here
the nonmagnetic SCF ground state of $H$, i.e.,

\begin{equation}
|\Phi_0>~=~\prod_{\nu\sigma,|\underline{k}|<k_F}~c^+_{\nu\sigma}(\underline{k})|0>.
\end{equation}

The strong correlations are taken intoaccount by including in the set
of $\{A_m(\underline{k})\}$ the Fourier transforms of a number of
local, on-site operators. They are 

\begin{equation}
A^{(1)}_{ij}(\ell)~  =  \left\{%
\begin{array}{r l}
2a^+_{i\uparrow}(\ell)\delta n_{i\downarrow}(\ell), & \mbox{$i=j$}\\[6pt] 
a^+_{i\uparrow}(\ell)\delta n_j(\ell),  & \mbox{$i\neq j$} 
\end{array} 
\right.~ 
\end{equation}
\begin{eqnarray*}
~A^{(2)}_{ij}(\ell)~ & = & \frac{1}{2}(a^+_{i\uparrow}(\ell)s^z_j(\ell)+a^+_{i\downarrow}(\ell)s^+_j(\ell))\\ \nonumber
~A^{(3)}_{ij}(\ell)~ & = & \frac{1}{2} a^+_{i\downarrow}(\ell)a^+_{j\uparrow}(\ell)a_{i\downarrow}(\ell).
\end{eqnarray*} 

The notation $\delta
n_{i\sigma}(\ell)=n_{i\sigma}(\ell)-<n_{i\sigma}(\ell)>$ and
$\underline{s}_i(\ell)=(1/2)\Sigma_{\alpha\beta}a^+_{i\alpha}(\ell)\underline{\sigma}_{\alpha\beta}a_{i\beta}(\ell)$
has been used here. The selected $A_m(\underline{k})$ consist therefore of
$A_{\nu}^{(0)}(\underline{k}) = c_{\nu \uparrow}^+ (\underline{k})$ and

\begin{eqnarray}
A_{ij}^{(r)}(\underline{k}) & = & \frac{1}{\sqrt{N}}\sum_{\ell} A_{ij}^{(r)}(\ell)e^{i\underline{k R}_{\ell}}\\ \nonumber 
~r~                         & = & 1, 2, 3. 
\end{eqnarray}

For a given value of $\underline{k}$ the total number of relevant
operators is 66. This requires the diagonalization of a 66 x 66 matrix
for each $\underline{k}$ point. In order to evaluate the matrix
elements we must first specify the wave operator $\Omega$. In
accordance with Eq. (2.11) we make the ansatz

\begin{equation}
|\Omega)=|1+\sum_{ij\ell}\eta_{ij}\delta O_{ij}(\ell))
\end{equation}

where the local operators $\delta O_{ij}(\ell)$ are given by

\begin{equation}
\delta O_{ij}(\ell)\;=\; \left\{%
\begin{array}{l l l}
2\delta n_{i\uparrow}(\ell)\delta n_{i\downarrow} (\ell), & i=j &\\[6pt]
\delta n_i(\ell)\delta n_j(\ell), & i\neq j &\\[6pt]  
\underline{s}_i(\ell)\underline{s}_j(\ell) & . 
\end{array}
\right. 
\end{equation}

\noindent The operators $\underline{s}_i(\ell)\underline{s}_j(\ell)$
generate Hund's rule correlations in the ground state of a transition
metal, while the other two types of operators reduce fluctuations of
charges from their average values, thus rendering the Coulomb
repulsions less effective. More details are found, e.g., in
Ref. [14]. The resulting spectrum for $Ni$ is shown in Fig. 6. All
parameter values are in units of the SCF bandwidth $W$. They are
obtained by fitting the measured multiplet structure of
transition-metal ions embedded in simple metals [84]. Shown is the SCF
density of states and the modifications which are obtained when the
correlations are included. In order to bring out more clearly the
changes caused by the different interactions we show in Fig. 6 c the
special case where the exchange constant $J$ as well as the
anisotropic exchange parameter $\Delta J$ have been set equal to
zero. When $J=\Delta J=0$ one obtains only one quasiparticle and one
satellite peak for each $\underline{k}$ point and given band index
$\nu$. The satellite peak in the density of states reflects the $e_g -
t_{2g}$ splitting caused by the ligand field. When $J$ is included,
but $\Delta J=0$ new spectral density appears near -2.1 and -0.7. This
reflects the atomic $d^2$ (hole) multiplet which splits into a $~^1S$
state, two degenerate singlets $~^1G$ and $~^1D$ and two degenerate
triplets $~^3P$ and $~^3F$. The splitting energy between $~^1S$ and
$~^1G$ is $5J$ while the one between $~^1G$ and $~^3F$ is $2J$. The
spacings between the structures at -2.1, -1.1 and -0.7 are of
comparable size. Inclusion of $\Delta J\neq 0$ splits the main
satellite at -1.1 into smaller peaks. We also show in Fig. 6c the
modifications in the spectrum which arise when we set $\Omega=1$,
i.e., when we neglect the correlations in $|\psi_0>$ and replace it by
the SCF ground state $|\Phi_0>$. In conclusion, we may state that the
correct position of the satellite peak as well as a band narrowing by
15 \% are obtained when experimentally determined parameters for the
interactions are used. The band narrowing is less than the observed
one which is approximately 25 \%.

The spectral density of $Ni$ can also be calculated by using Faddeev's
equations [85]. The results resemble the ones presented here but they
are not identical [86]. Faddeev's equations also contain
three-particle correlations which have not been included here. A
serious deficiency of the model is the neglect of the $4s$ band. This
might explain some discrepancies which arise when the present theory
is applied to other $3d$ transition metals. For example, we also
obtain multiplet structured satellites in the spectra of $Co$ and $Fe$
when we use the Coulomb parameters as determined from optical
experiments. However, such structures have not been observed in
photoemission experiments. Further details can be found in Ref. [83]. 

The long-term goal is certainly an extension of the theory to larger
basis sets. One would also like to take systematically into account
correlations $between$ sites, an extension not yet tried for the $3d$
transition metals. In principle, the projection technique is a very
suitable tool for modern computing.

\subsection{Spectral functions of $Cu-O$ planes}

The Hamiltonian for the electrons in the $Cu-O$ planes of cuprates has
been discussed before (see Eq. (3.7-9)). The main task for obtaining
the spectral densities is the proper choice of the set of relevant
operators $\{A_m(\underline{k})\}$.

The strong electron correlations, which must be accounted for,
determine the selection of these operators. They must include, first
of all, the hole operators
  
\begin{equation}
A_p(m,\underline{k})=p^+_{m\underline{k}\uparrow},~A_d(\underline{k})=d^+_{\underline{k}\uparrow}
\end{equation}

but also the Fourier transforms of a number of $local$ operators,
which generate the correlation hole surrounding the added particle or
hole. To those belong the Fourier transforms
$\bar{p}^+_{m\underline{k}\uparrow},\bar{d}^+_{\underline{k}\uparrow}$
of the local operators
$\bar{p}^+_{I\uparrow}=p^+_{I\uparrow}n_{p\downarrow}(I)$ and
$\bar{d}^+_{J\sigma}=d^+_{J\uparrow}n_{d\downarrow}(J)$,
respectively. They ensure reduced weights of configurations with
doubly occupied $Cu$ and $O$ orbitals, a consequence of the Coulomb
repulsions $U_d$ and $U_p$. Additional microscopic processes to be
included are defined by the operators [87]

\begin{eqnarray}
A_f(\underline{k})  & = & \frac{1}{\sqrt{N}}\sum_I~e^{-i\underline{kR}_I} p_{I\downarrow}^+S^+_I \nonumber \\    
A_a(\underline{k})  & = & \frac{1}{\sqrt{N}}\sum_I~e^{-i\underline{kR}_I} p_{I\uparrow}^+n_{d\downarrow}(I) \nonumber \\
A_c(\underline{k})  & = & \frac{1}{\sqrt{N}}\sum_I~e^{-i\underline{kR}_I} p_{I\uparrow}^+p_{I\downarrow}^+d_{I\downarrow}.
\end{eqnarray}

The $N$ different unit cells have been labeled by $I$ and their
lattice vectors by $\underline{R}_I$. The operator
$S_I^+=d_{I\uparrow}^+d_{I\downarrow}$ describes a spin flip of a hole
in the $3d_{x^2-y^2}$ orbital of site $I$. The operator
$p^+_{I\uparrow}$ refers to the combination 
 
\begin{equation}
p^+_{\uparrow} = \frac{1}{2}(p^+_{1\uparrow}-p^+_{2\uparrow}-p^+_{3\uparrow}-p^+_{4\uparrow}) 
\end{equation}

of the four $O$ orbitals surrounding $Cu$ site I. The operator
$A_f(\underline{k})$ ensures inclusion of processes, which result in
the formation of a singlet state between a hole on a $Cu$ site and
another one in a nearest-neighbor $O$ orbital (Zhang-Rice singlet
[88]). The operator $A_a(\underline{k})$ does the same for the triplet
state while $A_c(\underline{k})$ takes charge transfer processes in
the vicinity of the added particle (hole) into account. With this
choice of a total of nine relevant operators a 9 x 9 matrix
$G_{mn}(\underline{k},\omega)$ must be diagonalized for each
$\underline{k}$ point. Thereby the self-consistency condition (3.18)
is used throughout. The resulting densities of states for the case of
half-filling (corresponding to $La_2CuO_4$) and of hole doping (e.g.,
$La_{2-x}Sr_xCuO_4$) agree very well with those obtained from
numerical diagonalization of a cluster of four units of $CuO_2$
[89,90] (see Figs. 7,8). This proves the usefulness of the method. As
seen from those figures the half-filled system is insulating. The
structure around 2.5 $t_{pd}$ results from the singlet state, while
the one at $5t_{pd}$ represents the upper Hubbard band. When the
system is doped (see Fig. 8 b) spectral density is moved from the
upper Hubbard band to the region close to the Fermi energy. The system
is then metallic. In electron-doped systems like $Nd_{2-x}Ce_xCuO_4$
the Fermi level is in the upper Hubbard band. In that case spectral
density is shifted from the singlet states to the upper Hubbard band
which can accommodate an increasing number of electrons as the doping
increases. When the hole concentration approaches zero, the upper band
has a total spectral weight of 2 as compared with the half-filled case
where this weight is 1. For more details we refer to Ref. [87].

\setcounter{equation}{0}
\section{Electron Crystallization}

In metals the kinetic energy of the electrons is usually more
important than their mutual Coulomb repulsion. This is due to Pauli's
principle which results in Fermi energies of the order of a few
eV. However, this changes when the electron concentration is very
low. Let $r_0$ denote the mean radius of the volume an electron has
available, i.e., define $r_o$ via $(4\pi/3) r^3_o=\rho^{-1}$ where $\rho$ is
the electron density. The average kinetic energy $\delta\epsilon_{kin}$ of an
electron due to the uncertainty relation is
$\delta\epsilon_{kin}=(\Delta p)^2/2m\sim 1/(2m r^2_0)$. But the
average Coulomb repulsion is $\delta\epsilon_{pot}\sim e^2/r_0$ and
therefore larger than $\delta\epsilon_{kin}$ in the limit of large
$r_0$ or low densities. By considering a homogeneous electron gas with
the positive charge uniformly spread over the system (jellium model)
Wigner [20] discussed the form of the ground state in the low-density
limit and concluded that electrons crystallize in form of a lattice in
order to keep the Coulomb repulsion as low as possible. The kinetic
energy reduces to the zero-point motion of the electrons around their
equilibrium position. The change from a homogeneous to an
inhomogeneous electron charge distribution takes place at a value of
approximately $r_0/a_B\simeq 40 - 100$ where $a_B$ is the Bohr radius. At
those densities the electrons are sufficiently far apart that the
exchange plays only a minor role because of its exponential decrease
with distance. The excitations are given by the vibrations of the
electrons around their lattice positions and the low-temperature
specific heat $C\sim T^3$ like for a phononic system. Possible
realizations of Wigner crystal are found in semiconducting inversion
layers [91, 92], or rare-earth pnictides with low carrier concentrations
[93]. For a comprehensive review see Ref. [94]. Often is is more
appropriate to consider electrons on a lattice. In order to explain
the dramatic temperature dependence of the resistivity of magnetite
($Fe_3 O_4$), Verwey [95] developed a model for charge ordering in
that rather complex spinel structure. According to his model the $2 \cdot
Fe^{3+}+1 \cdot Fe^{2+}$ ions per formula unit are distributed at low
temperatures as follows. One $Fe^{3+}$ is used to form a stable cubic
sublattice. The remaining $Fe^{3+}$ and the $Fe^{2+}$ ions form a structure
on which they alternate. This is achieved by assuming that the
$Fe^{3+}-Fe^{3+}$ nearest-neighbor interaction energy is higher than the
$Fe^{2+}-Fe^{3+}$ one, so that $Fe^{3+}$ ions prefer $Fe^{2+}$ ions
on nearest neighbor sites. The kinetic energy of the electrons is
discarded in Verwey's theory. According to this model a first-order phase
transition of order-disorder type takes place at high
temperatures. The ground state suggested by Verwey can be considered
as a form of electron crystallization which is quite distinct from the
one considered by Wigner. Its explicit form has been questioned though,
by neutron-scattering experiments. 

A third form of
electron crystallization is due to Mott [96,97] and Hubbard [98,69],
and is also based on electrons positioned on a lattice. Here it is the
on-site Coulomb repulsion of electrons which may lead to
crystallization, provided the system is at half-filling (i.e., with
one electron per site), and the Coulomb repulsion is sufficiently
large as compared with the hopping matrix element between sites. Mott
realized that this is always the case if a chain of $H$ atoms is
considered and the latter are pulled sufficiently apart. Hubbard
discussed the metal-insulator transition associated with
electron-crystallization by suggesting various approximations for the
computation of the excitation spectrum of the Hamiltonians (3.1) at
half filling. 

Recently it has become clear that still another modification of
electron crystallization is realized in $Yb_4 As_3$
[30,22,99,100]. There have been previous observations on charge
ordering on $Yb_4 As_3$ and $Sm_4 Bi_3$, without interpretations
offered [21, 101]. Like in the case of the Wigner lattice it is here
the long-range Coulomb interaction which results in what is actually a
crystallization of holes. But in distinction to the Wigner case
crystallization takes place at high densities. This is so since $4f$
holes are involved here which have a very small kinetic energy due to
the small hybridization of the well localized $4f$ orbitals.

$Yb_4As_3$ is of the anti-$Th_3P_4$ structure. The $Yb$ ions are positioned
on four families of interpenetrating chains which point along the four
diagonals of a cube. It is important that the distance of neighboring
$Yb$ ions within a chain is 3.80 {\AA} and therefore larger than of $Yb$
ions between different chains which is 3.40 {\AA}. The three nearest
neighbors of a $Yb$ ion therefore belong to the other three families
of chains. The structure is shown in Fig. 9.

We write $Yb_3^{2+}Yb^{3+}As_3$ in order to demonstrate that the
system has one $4f$ hole $(Yb^{3+}\Longrightarrow 4f^{13})$ per
formula unit. At high temperatures the $4f$ holes are delocalized and
the system is metallic. Measurements of the Hall coefficient show that
the carrier concentration is indeed approximately 1/4 per Yb ion. At
$T_S=300~K$ the system undergoes a weak first-order phase transition
below which the $Yb^{3+}$ ions accumulate on one family of chains,
e.g., those along the [111] direction. Since the $Yb^{3+}$ ions are
smaller than the $Yb^{2+}$ ions, the phase transition is accompanied
by a volume conserving trigonal distortion. Thereby chains in the [111]
direction are shortened while those in the other three directions parallel to
the diagonals of a cube are elongated. 

The driving mechanism of the phase transition is the Coulomb repulsion
of holes. The energy is minimized when the holes move into one family
of chains because of the large distances between ions in a chain.
Disregarding first the structural changes associated with the phase
transition we are faced with a Hamiltonian for the holes of the form

\begin{equation}
H= -t {\sum_{ij\sigma}}' \hat{a}^+_{i\sigma}\hat{a}_{j\sigma}+
{e^2\over 2} 
\sum_{ij} \frac{e^{-\lambda R_{ij}}}
{R_{ij}} \hat{n}_i\hat{n}_j + \sum_{<ij>} J_{ij}\underline{S}_i\underline{S}_j.
\end{equation}

The operators $\hat{a}_{i\sigma}^+$, $\hat{a}_{j\sigma}$ are the same
as in Eq. (3.3). The first term describes the kinetic energy of the
$4f$ holes and the prime indicates that nearest neighbors are
considered only. The second term describes the screened Coulomb interaction. We
denote with $R_{ij}=|\underline{R}_i-\underline{R}_j|$ the distance
between lattice sites $i$ and $j$ and $\lambda^{-1}$ is the screening
length which depends on the carrier concentration. The last term is
due to the exchange and $<ij>$ refers of pairs of ions in the same
chain. While exchange plays a minor role in a Wigner
crystal due to the large lattice constant it is of importance when
crystallization takes place at high densities as it is the case here. 

Considering only the Coulomb term without taking screening into
account except by using an effective charge, it has been demonstrated that the
energy difference between a uniform distribution of holes and one with the
holes concentrated in one family of chains is of the order of a few $meV$ per
formula unit [102]. A comparison with the effective $f$ bandwidth of
order 0.2 eV makes plausible that hole crystallization will take
place. Note that there are also other charge-ordered configurations
of low energy. For example, an ordering of charges in the form
$Yb^{3+}-Yb^{2+}-Yb^{2+}-Yb^{2+}-Yb^{3+}...$ in all four families of
chains has also a low Coulomb-repulsion energy. In fact, von Schnering
and Grin have found that the Coulomb-repulsion energy is slightly
lower than the one with all $Yb^{3+}$ placed into one family of
chains. However, that state is fourfold degenerate and therefore can
further lower its energy by a Jahn-Teller distortion. This leads back
to a state with short and long chains and therefore with holes
concentrating in the short chains [30]. However, that order will not
be a perfect one for the reason that the $4$-fold degenerate state has
a slightly lower repulsion energy. Thus, the optimal state must have
incomplete order. Experimentally a 10 \% deviation from perfect order
is observed [22]. A different way of looking at imperfect charge order
is to realize that the zero-point fluctuations of the holes on chains
in the [111] direction necessarily leads to a spreading of holes into the
long chains. For a rough estimate of the effect see Ref. [100].

The above physical picture justifies a description of the phase
transition in terms of a band Jahn-Teller effect [30]. The
corresponding Hamiltonian is 

\begin{equation}
H = -t \sum^4_{\mu=1}~\sum_{<ij>\sigma}(a^+_{\mu i\sigma}a_{\mu j\sigma} + h.c.)
-\epsilon_{\Gamma}\sum^4_{\mu=1}
\sum_{i\sigma}\Delta_{\mu} n_{\mu i\sigma}+c_{\Gamma}^{(0)}\epsilon^2_{\Gamma}N_0.
\end{equation}

Here $\mu$ is a chain index and $i$ labels a site on chain
$\mu$. Furthermore, $\epsilon_{\Gamma}(\Gamma=\Gamma_5)$ is a volume
conserving strain order parameter which couples to the deformation
potential $\Delta_{\mu}=\Delta[(\delta_{\mu 1}-(1-\delta_{\mu 1})/3]$
and $c_{\Gamma}^{(0)}$ is the associated elastic constant. $N_0$ is
the number of sites. A Jahn-Teller phase transition takes place if
$\Delta^2/(t c^{(0)}_{\Gamma})>3$. From LDA calculations one can
estimate $4t\approx0.2 eV$. Furthermore,
$c^{(0)}_{\Gamma}/\Omega=4\cdot 10^{11}$ erg/$cm^3$ where $\Omega$ is
the volume of the unit cell. We choose $\Delta=5 eV$ which gives a
Gr\"uneisen parameter of $\Omega_c=\Delta/(4t)\simeq~25$ typical for
intermediate valence compounds. With this set of parameters a
transition temperature $T_S\simeq 250 K$ is obtained in approximate
agreement with the observed value [30]. Note that strong correlations
not included in (4.2) may influence considerably the details of the
transition [103]. As discussed in Sect. 5.3 the low-temperature
specific heat of $Yb_4As_3$ is of the form $C=\gamma T$ and therefore
quite different from that of a Wigner crystal. This is due to the
exchange interactions which are important in the crystalline
phase. The charge ordering leads to quasi one-dimensional Heisenberg
spin chains and it is known that the latter have a specific heat linear in $T$.

$Yb_4As_3$ is most likely not the only material showing electron or
hole crystallization. Other candidates are $Eu_4As_3$ and $Eu_3S_4$
where M\"ossbauer and other measurements have shown that at low
temperatures the $Eu^{3+}$ and $Eu^{2+}$ ions are at fixed lattice
sites [130, 131]. More precisely, the valence fluctuation times must
be larger than $\tau=10^{-8}$ sec, a typical testing time in a
M\"ossbauer experiment. It should be mentioned that there has been an
attempt to explain the experiments on $Eu_4As_3$ and $Eu_3 S_4$ by a
Verwey transition [132 - 134].

\setcounter{equation}{0}
\section{HEAVY FERMIONS}

The investigation of metallic systems with heavy quasiparticle
excitations has developed into an own branch of low-temperature
physics. In most cases these systems contain $Ce, Yb, U$ or $Np$ ions
as one of their constituents, implying that $4f$ or $5f$ electrons are
essential. Examples are the metals $Ce Al_3$, $CeCu_2 Si_2$, $Ce Ru_2
Si_2$, $CeCu_6$, $Yb Cu_2 Si_2$, $U Be_{13}$, $U Pt_3$, and $Np
Be_{13}$. For experimental reviews see Refs. [23,24] and [26,28],
respectively. But also the electron-doped cuprate $Nd_{2-x}Ce_x Cu
O_4$ shows heavy-fermion behavior [32] in the range $0.1\leq x \leq
0.2$. Heavy quasiparticles have also been found in semimetals like
$Yb_4 As_3,Sm_3S_4$ or in some of the monopnictides and even in
insulators like $Yb B_{12}$ or $Sm B_6$ [104].

The following experimental findings define a heavy-fermion system:\\
(a) \begin{minipage}[t]{15cm} 
A low temperature specific heat $C=\gamma T$ with a $\gamma$
coefficient of order $1J mol^{-1}K^{-2}$, rather than $1 m J mol^{-1}
K^{-2}$ as, e.g., found for sodium metal;
\end{minipage}\\[3ex]
(b) A Pauli spin susceptibility $\chi_S$ which is similarly enhanced
as $\gamma$;\\ 
(c) A ratio $R=\pi^2k_B^2~\chi_S/(3\mu^2_{eff}\gamma)$
(Sommerfeld-Wilson ratio) of order unity.\\ 

\noindent Both quantities, $\gamma$ and $\chi_S$ are proportional to
the quasiparticle density of states $N^*(0)$ per spin direction at the
Fermi level. The latter is proportional to $m^*$, the effective mass
of the quasiparticles. When $R$ is calculated the density of states
$N^*(0)$ drops out. For free electrons $R=1$, while in the presence of
quasiparticle interactions $R=(1+F_0^a)^{-1}$ where $F^a_0$ is a
Landau parameter. When conditions (a)-(c) are met, one may assume a
one-to-one correspondence between the low-energy excitations of the
(complex) system like $CeAl_3$ and those of a free electron gas,
provided a strongly renormalized effective mass $m^*$ is used and, in
the case of semimetals or insulators, an effective charge $e^*$,
instead of the corresponding bare quantities. 

Heavy-fermion behavior requires the presence of a characteristic
low-energy scale in the system. The latter is usually characterized by
a temperature $T^*$. As the temperature $T$ of the system exceeds $T^*$
the quasiparticles lose their heavy-mass character. The specific heat
levels off, and the spin susceptibility changes from Pauli- to
Curie-like behavior. With further increase of temperature the
rare-earth or actinide ions behave more and more like ions with
well-localized $f$ electrons.

A key problem is to understand the physical origin of the low-energy
excitations. For a long time it was believed that the Kondo effect is
the sole source of heavy quasiparticles. The physics of the Kondo
effect is extensively described in a monograph [28] and a number of
reviews [23-27]. However, by now it is known that also other effects
may lead to heavy-fermion behavior. In all cases a lattice of $4f$ or
$5f$ ions is involved though. In metallic systems this lattice couples
to the conduction electrons. The latter are either weakly correlated
like in $Ce Al_3$ or strongly correlated like in the cuprates which
may become high-$T_c$ superconductors. Strong correlations among the
conduction electrons may influence substantially the physical
properties of the system. Such a situation is encountered in
$Nd_{2-x}Ce_xCu O_4$ and as shown below it is here the Zeeman effect
which is responsible for the formation of heavy-fermion
excitations. In the semimetal $Yb_4 As_3$ the heavy quasiparticles are
intimately related to quasi one-dimensional chains of $Yb^{3+}$ ions
which interact antiferromagnetically with each other. It is well known
that a Heisenberg chain has a linear specific heat $C=\gamma T$ at low
temperatures and a Pauli like susceptibility. Thus, instead of having
a single physical origin, heavy fermions may result from a variety of
effects. 

Obviously, the low lying excitations, the main feature of
heavy-fermion systems involve predominantly spin degrees of
freedom. Evidential is the entropy $S$ associated with the excess
specific heat. It is of order $S\simeq k_B\ell n\nu_f$ per $f$ site,
where $\nu_f$ is the degeneracy of the crystal-field ground state of
the incomplete atomic $f$ shell. In the following we discuss the three
different routes to heavy-fermion behavior just outlined. It is likely
that they will be supplemented by other ones in the future.  

\subsection{Kondo lattices}

The essence of the single-site Kondo effect is the formation of a
singlet ground state due to a weak hybridization of the incomplete
$4f$ shell with the conduction electrons. We derive the singlet
wavefunction by starting from the Anderson impurity Hamiltonian 

\begin{eqnarray}
H & = & \sum_{km}\epsilon(k)c^+_{km}c_{km}+\epsilon_f\sum_m n^f_m 
+ \frac{U}{2}\sum_{m\neq m'}~n_m^f n_{m'}^f+ \\ \nonumber
~ & + &  \sum_{km}V(k)(f^+_mc_{km}+c^+_{km}f_m)~+\tilde{H}_0.
\end{eqnarray}

Here $f^+_m$ denotes the creation operator of an $f$ electron in state
$m$ of the lowest $J$ multiplet and $n^f_m = f^+_mf_m$. The
$f$-orbital energy is $\epsilon_f$ and $U$ is the $f-f$ Coulomb
repulsion. The $c^+_{km}$ create conduction electrons with momentum
$|\underline{k}|=k$ and the quantum numbers $\ell = 3, J$ and $m$. The
hybridization between the $f$ and conduction electrons is given by the
matrix element $V(k)$. Finally, $\tilde{H}_0$ contains all those
degrees of freedom of the conduction electrons which do not couple to
the $4f$ shell. The following ansatz for the singlet ground-state wave
function is due to Varma and Yafet [105]

\begin{equation}
|\psi_0> = A(1+\frac{1}{\sqrt{\nu_f}}\sum_{km}\alpha(k)f^+_m c_{km})|\phi_0>
\end{equation}

where $|\phi_0>$ represents the Fermi sea of the conduction electrons.
It is closely related to the one suggested by Yoshida [106] for the
ground state of the Kondo Hamiltonian. The variational parameters $A$
and $A\alpha(k)$ are obtained by minimizing the energy
$E_0=<\psi_0|H|\psi_0> / <\psi_0|\psi_0>$. The latter is always lower
than the one of the multiplet $\mid\psi_m>=f^+_m|\phi_0>$. The
difference $\epsilon$ is found to be

\begin{equation}
\epsilon = - D~exp[-|\epsilon_f|/(\nu_f N(0)V^2)]
\end{equation}

and denotes the energy gain due to the formation of the singlet. Here
$D$ is half of the bandwidth of the conduction electrons and $N(0)$
is their density of states per spin direction. It is customary to
associate a temperature $T_K$, the Kondo temperature with this energy
gain. The singlet-triplet excitation energy $-\epsilon$ is often of
the order of a few meV only, and provides a low-energy scale. When a
lattice of $f$ sites is considered instead of a single one, e.g., like
in $Ce Al_3$ the Anderson-lattice Hamiltonian is replacing
Eq. (5.1). The energy scale $k_BT_K$ is then replaced by a related
one, $k_BT^*$, which includes modifications due to the interactions
between different $f$ sites. The energy gain due to the formation of
singlets competes with the one of magnetic $f$-sites interacting via
the RKKY interaction [107]. The latter always wins for small enough
hybridization $V$, because it is proportional to $V^4$ whereas
$k_BT^*$ depends exponentially on $V$ (see Eq. (5.3)) and therefore is
smaller in that limit. This seems to be the case in systems like
$CeAl_2$, $CePb_3$ and $Np Be_{13}$ which are antiferromagnets at low
temperatures. 

In addition to $T^*$ there does exist another characteristic
temperature $T_{coh}<T^*$ below which the local singlet-triplet
excitations lock together and form coherent quasiparticle excitations
with large effective mass $m^*$. The details of this transition are
still an open problem, but de Haas-van Alphen measurements have
demonstrated convincingly that at low temperatures the $f$ electrons
behave like delocalized electrons [108]. They contribute to the Fermi
surface and to large effective mass anisotropies. It is surprising
that one can calculate the Fermi surface of some of the heavy-fermion
systems and determine the mass anisotropies with one adjustable
parameter only. This is achieved by renormalized band-structure
calculations (for reviews see [109, 110]). They are based on a
description of the effective potential seen by a quasiparticle in
terms of energy-dependent phase-shifts $\eta^A_{\ell}(\epsilon)$ of
the different atoms $A$. The index $\ell$ refers here to the different
angular momentum channels. As an example we discuss in the following
the calculation of the Fermi surface of $Ce Ru_2 Si_2$ [111]. The
essential approximation is to use for the phase shifts the ones
computed within the LDA, with the exception of the $\ell=3$ phase
shift of the $Ce$ ions. Thus, only the $\eta^{Ce}_{\ell=3}(\epsilon)$
phase shift remains undetermined. It contains the strong correlations
of the $4f$ electrons and cannot be properly evaluated within the
LDA. This approximation neglects virtual transitions between different
crystal-field eigenstates caused by the coupling between conduction
and $4f$ electrons. (The mass enhancement of the conduction electrons
in $Pr$ metal falls into that category [112]).

According to Hund's rules the ground state multiplet of $Ce^{3+}$ with
a $4f^1$ configuration is $j=5/2$. The multiplet $j=7/2$ is
sufficiently high in energy that it may be neglected and therefore
$\eta^{Ce}_{j=7/2}(\epsilon_F)=0$. Of the $j=5/2$ multiplet, only the
Kramers degenerate crystal-field ground state is taken into account,
because it is the only one occupied at low temperatures. Therfore,
near the Fermi energy only the phase shift function
$\eta_{\tau}^{Ce}(\epsilon)(\tau=1,2)$ among the different $\ell=3$
channels differs from zero.  It contains the strong electron
correlations and its form is unknown. In the spirit of Landau's
Fermi-liquid theory we expand this function in the vicinity of
$\epsilon_F$ and write 

\begin{equation}
\eta_{\tau}^{Ce}(\epsilon) = \eta_{\tau}^{Ce}(\epsilon_F)+a(\epsilon-\epsilon_F)+O((\epsilon-\epsilon_F)^2).
\end{equation}

The expansion contains the two unknown parameters
$\eta_{\tau}^{Ce}(\epsilon_F)$ and $a$. One of them, i.e.,
$\eta_{\tau}^{Ce}(\epsilon_F)$ is fixed by the requirement that a $Ce$
site contains one $4f$ electron ($n_f=1$). According to Friedel's sum
rule this implies $\eta_{\tau}^{Ce}(\epsilon_F)=\frac{\pi}{2}$. The
remaining parameter $a$ fixes the slope of the phase shift at
$\epsilon_F$. It therefore determines the density of states and with
it the effective mass of the quasiparticles. We set $a =
(k_BT^*)^{-1}$ and determine $T^*$ by the requirement that the
specific heat coefficient $\gamma$ calculated from the resulting
quasiparticle dispersion agrees with the experimental
one. Calculations of this form have explained and partially predicted
[109,111] the Fermi surface and the large mass anisotropies of $Ce
Ru_2 Si_2$ (see Table 10 and also Fig. 10) [113, 114]. For more details
on renormalized band theory we refer to the comprehensive reviews
[109, 110].

When the temperature exceeds $T_{coh}$ the excitations lose their
coherence properties and we are dealing with approximately independent
scatterers. In that regime the specific heat contains large
contributions from the incoherent part of the $f$ electron excitations.

The noncrossing approximation (NCA) is a valuable tool for treating
the coupled $4f$ and conduction electrons in that temperature regime
[116-118]. It leads to a system of coupled equations of the form

\begin{eqnarray}
\Sigma_0(z) & = & \frac{\Gamma}{\pi}\sum_m~\int\limits^{+\infty}_{-\infty}d\varsigma\rho_m(\varsigma)K_+(z-\varsigma)\\ \nonumber
\Sigma_m(z) & = & \frac{\Gamma}{\pi}\int^{+\infty}_{-\infty} d\varsigma\rho_0(\varsigma)K_--(z-\varsigma). 
\end{eqnarray} 

Here $\Gamma = \pi N(0)V^2$ and $K_{\pm}(z)$ are defined by

\begin{equation}
K_{\pm}(z)=\frac{1}{N(0)}\int\limits^{+\infty}_{-\infty}~d\epsilon \frac{N(\pm\epsilon)f(\epsilon)}{z+\epsilon}
\end{equation} 

where $f(\epsilon)$ is the Fermi energy and $N(\epsilon)$ is the
energy-dependent conduction-electron density of states. The function
$\Sigma_{\alpha}(z)$ and $\rho_{\alpha}(z)(\alpha = 0, m)$ relate to
each other through

\begin{eqnarray}
\rho_{\alpha}(z) & = & -\frac{1}{\pi} Im\{R_{\alpha}(z)\}\\ \nonumber
R_{\alpha}(z)    & = & \frac{1}{z-\epsilon_{\alpha}-\Sigma_{\alpha}(z)}
\end{eqnarray}

with $\epsilon_{\alpha=0}=0$, $\epsilon_{\alpha=m}=\epsilon_{fm}$. The
NCA equations have to be solved numerically [119]. However, one can
find simple, approximate analytic solutions which have the advantage
that crystal-field splittings can be explicitly included, a goal which
has not been achieved yet by numerical methods. Once the
$\rho_{\alpha}(\epsilon)$ are known, one can determine, e.g., the
temperature dependence of the $f$-electron occupancies $n_{fm}=<f^+_m
f_m>$ through 

\begin{equation}
n_{fm}(T)=\frac{1}{Z_f}\int\limits^{+\infty}_{-\infty}d\epsilon\rho_m(\epsilon)e^{-\beta(\epsilon-\mu)},
\end{equation}

where $\mu$ is the chemical potential and

\begin{equation}
Z_f=\int_C \frac{dz}{2\pi i} e^{-\beta z}(R_0(z)+\sum_m R_m(z))
\end{equation}

is the partition function of the $f$ electrons. Knowing the
$n_{fm}(T)$ one can compute quantities like the temperature dependence
of the quadrupole moment of the $f$ sites

\begin{equation}
Q(T)=\sum_m <m|(3J^2_z-J^2)|m>n_{fm}(T).
\end{equation}

The theory has been used to explain the observed $Q(T)$ behavior of
$Xb$ in $Yb Cu_2 Si_2$ [120, 121].

When $T\gg T^*$, the $f$ electrons can be treated as being
localized. Via an exchange coupling they are weakly interacting with
the conduction electrons and perturbation theory can be applied in
order to study the resulting effects. 

A beautiful justification of the above scenario is the observed
difference in the Fermi surface of $CeRu_2Si_2$ and $CeRu_2Ge_2$ which
is shown in Fig. 10. When $Si$ is replaced by $Ge$ the distance
between $Ce$ and its nearest neighbors is increased. This causes a
decrease in the hybridization of the $4f$ electrons with the valence
electrons of the neighboring sites. While in $CeRu_2Si_2$ the
characteristic temperature is $T^*\simeq 15 K$, it is practically zero
in $CeRu_2Si_2$. De Haas-van Alphen experiments are performed at a
temperature $T\simeq 1K$ implying that for $CeRu_2Si_2$ it is $T\ll
T^*$ while for $Ce Ru_2Ge_2$ one is in the regime $T\gg
T^*$. Therefore, the $4f$ electron of $Ce$ contributes to the volume
enclosed by the Fermi surface of $CeRu_2 Si_2$, but not of $CeRu_2
Ge_2$. Indeed, Fig. 10 shows that the two Fermi surfaces have similar
features, but the enclosed volumes differ by one electron. The Fermi
surface of $CeRu_2Ge_2$ has a decreased electronic part and an
increased hole part as compared with the one of $CeRu_2Si_2$.

\subsection{Zeeman scenario - $Nd_{2-x}Ce_xCu O_4$}

Low-temperature measurements of the specific heat and magnetic
susceptibility have demonstrated the existence of heavy quasiparticles
in the electron doped cuprate $Nd_{2-x}Ce_xCu O_4$ [32]. For $x=0.2$
and temperatures $T \leq 1K$ the linear specific-heat coefficient
is $\gamma=4J/(mol \cdot K^2)$. The magnetic susceptibility $\chi_s$ is
approximately $T$-independent in that temperature regime and the
Sommerfeld-Wilson ratio is $R\simeq 1.8$. The experimental findings
are shown in Fig. 11. While these features agree with those of other
heavy-fermion systems, there are also pronounced differences. In
superconducting heavy-fermion systems like $CeCu_2 Si_2$ or $U Pt_3$
the Cooper pairs are formed by the heavy quasiparticles. This is
evidenced by the fact that the jump in the specific heat $\Delta C$ at
the superconducting transition temperature $T_c$ is directly related
to the large $\gamma$ coefficient, i.e., $\Delta C(T_c)/(\gamma
T_c)\approx 2.4$. The low-energy excitations are therefore strongly 
reduced below $T_c$ because one must overcome the binding energy of
the pairs. But in superconducting $Nd_{1.85}Ce_{0.15}CuO_4$ the
formation of Cooper pairs has no noticeable effect on the
heavy-fermion excitations. They remain uneffected by superconductivity.

A crucial difference between $Nd_{2-x}Ce_xCuO_4$ and,
e.g. $CeCu_2Si_2$ are the strong electron correlations between the
conduction electrons present in the former, but not in the latter
material. In the two-dimensional $Cu-O$ planes of $Nd_{2-x}Ce_xCu O_4$
with $x \geq 0.1$ we have to account for antiferromagnetic fluctuations. 
There is considerable experimental evidence that these fluctuations
are very slow at low temperatures. Consider undoped $Nd_2CuO_4$, an
antiferromagnet with a N$\acute{e}$el temperature of $T_N \simeq
270~K$. The exchange interactions between a $Nd$ ion and its
nearest-neighbor $Cu$ ions cancel because of the antiferromagnetic
alignment of the $Cu$ spins. Therefore one is left with the
next-nearest neighbor $Cu-Nd$ spin interaction. It is of the form
$\alpha\underline{s}_{Cu}\underline{S}_{Nd}$ and larger than the $Nd-Nd$
interaction. The Schottky peak in the specific heat seen in Fig. 11
results from the spin flips of the $Nd$ ions in the staggered
effective field $\alpha< \underline{s}_{Cu}>$ set up by the $Cu$ spins
(Zeemann effect). It is also present in doped systems like
$Nd_{1.8}Ce_{0.2}CuO_4$ where antiferromagnetic long-range order is
destroyed by doping. This can only be understood if the changes in the
preferred direction of the $Cu$ spins occur sufficiently slowly,
i.e., slower than $10^{-10}$ sec in the present case, to that the $Nd$
spins can follow those changes adiabatically. Only then is a similar
energy so that in $Nd_2CuO_4$ required to flip a $Nd$ spin. This
physical picture has been confirmed by recent inelastic
neutron-scattering and $\mu SR$ experiments [122, 123]. Spin-glass
behaviour can be excluded.  

When $Nd$ ions are replaced by $Ce$ ions, the latter contribute
approximately 0.5 electrons more to the $Cu-O$ planes than the
former. Thus a corresponding number of $Cu$ sites are in a $3d^{10}$
configuration. These sites have no spin and consequently they do not
interact with the $Nd$ ions. The extra electrons move freely in the
$Cu-O$ planes and therefore, the interaction of a $Nd$ ion with the
next-nearest $Cu$ site is repeatedly turned off and on. It is this
feature which results in heavy-quasiparticles. 

Two model descriptions have been advanced in order to explain the
low-energy excitations of $Nd_{2-x}Ce_xCuO_4$. One is based on a
Hamiltonian in which the $Nd-Cu$ interaction is treated by a
hybridization between the $Nd\ 4f$ and $Cu\ 3d$ orbitals. Usually it is
much easier to extract heavy quasiparticles from such a Hamiltonian
than from one with a spin-spin interaction like the Kondo
Hamiltonian. The slow, antiferromagnetic fluctuations of the $Cu$
spins are replaced by a static staggered field acting on them. This
symmetry-breaking field also accounts for the strong correlations in
the $Cu-O$ planes because charge fluctuations between $Cu$ sites are
strongly reduced this way (unrestriced Hartree-Fock). The Hamiltonian
$H$ reads therefore 

\begin{eqnarray}
H & = & - t\sum_{<ij>\sigma}(a^+_{i\sigma}a_{j\sigma}+h.c.)+h\sum_{i\sigma} \sigma e^{i\underline{Q R}_i}a^+_{i\sigma}a_{i\sigma}\\ \nonumber
~ & + & V \sum_{i\sigma} (a^+_{i\sigma}f_{i\sigma}+h.c.) + \tilde{\epsilon}_f\sum_{i\sigma}f^+_{i\sigma}f_{i\sigma}. 
\end{eqnarray}

Here $\underline{Q}=(\pi,\pi)$ is a reciprocal lattice vector,
$\underline{R}_i$ denotes the positions of the $Cu$ ions and $h$ is
the staggered field. The operators $a^+_{i\sigma},f^+_{i\sigma}$
create an electron in the $Cu\ 3d_{x^2-y^2}$ and $Nd\ 4f$ orbital,
respectively. For simplicity, only one $Nd$ site per $Cu$ site is
considered and one $4f$ orbital with energy $\tilde{\epsilon}_f$ is
assumed instead of seven. The energies $\tilde{\epsilon}_f$ and $V$
are strongly renormalized quantities because of the $4f$ electron correlations.

The Hamiltonian (5.11) is easily diagonalized. Four bands are
obtained, two of which are $d$-like ($Cu$) and two which are $f$-like
(Nd). The dispersions of the four bands are given by 

\begin{equation}
E_{\nu}(\underline{k}) = \frac{\tilde{\epsilon}_f\pm\epsilon_{\underline{k}}}{2}\pm\frac{1}{2}
\sqrt{(\epsilon_{\underline{k}}\mp\tilde{\epsilon}_f)^2+4V^2},~~\nu=1,...,4
\end{equation}

where
$\epsilon_{\underline{k}}=(\epsilon^2_0(\underline{k})+h^2/4)^{\frac{1}{2}}$
and $\epsilon_0(\underline{k})=-2t(cos k_x + cos k_y)$. They are
schematicly shown in Fig. 12. At half-filling only the lower $f$ band
is filled and the Schottky-peak contributions to $C(T)$ are due to
transition from the filled lower to the empty upper $f$ band. When the
planes are doped with electrons the upper $f$ band becomes partially
filled resulting in low-energy intraband excitations with large
effective mass. The latter follows from the quasiparticle dispersion 

\begin{equation}
E_{qp}(\underline{k})\simeq \tilde{\epsilon}_f + \frac{V^2}{(\tilde{\epsilon}_f+\epsilon(\underline{k}))}.
\end{equation}

In the present case it is the Zeeman splitting of the $f$ states which
is responsible for the occurrence of heavy-electron behaviour. The
effect of superconductivity on the heavy quasiparticles can be studied
by adding an attractive part for the charge carriers in the $Cu-O$
planes to the Hamiltonian. The latter can again be diagonalized. When
the density of states is calculated one finds unchanged contributions
from the $f$ bands inside the BCS gap. They originate from the $Nd$
spin degrees of freedom and explain why the heavy quasiparticles
remain unaffected by superconductivity.

The second model description of the $Nd$ spins coupled to the $Cu$
spin is based on stochastic forces acting on the latter [124]. They
mimic the interaction of a $Cu$ spin with the other $Cu$ spins. In
that case we start from the Hamiltonian 

\begin{equation}
H_{int}=\alpha \underline{s}_{Cu}\cdot\underline{S}_f~~~,~~~\alpha>0
\end{equation}

describing the $Nd-Cu$ interaction. For simplicity, both spins are assumed to
be of magnitude $S$. We treat the vector
$\underline{\Omega}=\underline{s}_{Cu}/S$ like a classical variable,
subject to a stochastic force. We assume a Gauss-Markov process in
which case the distribution function obeys a Fokker-Planck
equation. The correlation function is then of the form 

\begin{equation}
<\underline{\Omega}(0)\underline{\Omega}(t)>~=~e^{-2 D_rt}
\end{equation}

where $D_r$ can be obtained from the nonlinear $\sigma$ model
[125]. Because there is no long-range order
$<\underline{\Omega}(t)>=0$. The motion of the $Nd$ spin is governed
by the equation 

\begin{equation}
\frac{d}{dt}~\underline{n}(t)~=~\omega_0(\underline{\Omega}(t)\times \underline{n}(t))
\end{equation}

where $\underline{n}(t)=\underline{S}_f/S$ and $\omega_0=\alpha
S$. The spectral function 

\begin{equation}
I(\omega) = \frac{1}{2\pi} \int\limits^{+\infty}_{-\infty} dt e^{i\omega t}<\underline{n}(0)n(t)>
\end{equation}

is evaluated by making use of the corresponding stochastic Liouville equation.
We find that $I(\omega)$ is of the form

\begin{equation}
I(\omega) = \frac{1}{3\pi}\frac{4D_r}{\omega^2+(4D_r)^2}+~~({\rm
side~peaks~at}\ \omega_0).
\end{equation}

While $D_r(T)$ vanishes as $T\to 0$ in the presence of long-range
order, it remains finite when the latter is destroyed by doping. A
linear specific-heat contribution of the $4f$ spin is obtained from

\begin{equation}
C(T)_{imp}=\frac{d}{dT}<H_{int}>=\frac{S(S+1)}{T^2}\int\limits^{\infty}_0 d\omega\omega^2\frac{I(\omega)}{cos h^2(\omega/2T)}
\end{equation}

when $D_r(T=0)\neq0$. The side peaks of $I(\omega)$ give raise to a
Schottky-type contribution. The calculated specific heat reproduces
the experiments reasonably well. One shortcoming of the theory in its
present form is the low-temperature spin susceptibility which follows
from 

\begin{equation}
\chi_{imp}(T)=\frac{4}{3}(g\mu_B)^2 S(S+1)\int\limits^{\infty}_0 d\omega \frac{I(\omega)}{\omega}tan h \frac{\omega}{2T}.
\end{equation}

Instead of a temperature independent Pauli paramagnetism we find
$\chi_{imp}(T)\sim \ell n(D_r/T)$ at low $T$. This is possibly due to the
neglect of $Nd-Nd$ interactions. However, when evaluated for $T=0.4~K$
one obtains for $Nd_{1.8}Ce_{0.2}Cu O_4$ a Sommerfeld-Wilson ratio of
$R\simeq 1.4$. 

\subsection{Hubbard route: $Yb_4As_3$}

The semimetal $Yb_4 As_3$ is an example of a system in which $4f$
holes crystallize at low temperatures. The structure of the material
and the transition from a metallic high-temperature phase to a
semimetallic low-temperature phase were discussed in Sect. IV. Here we
concentrate on the heavy-fermion properties at low temperatures. The
following experimental observations are relevant in that
respect. Measurements of the Hall constant demonstrate that at low
temperatures only one carrier per $10^3\ Yb$ ions remains. We
interprete this being due to the $Yb^{3+}$ chains representing almost
half-filled Hubbard systems. The resistivity is at low temperatures of
the form $\rho(T)=\rho_0+AT^2$ like for a Fermi liquid. A linear specific
heat is found at low $T$ with a $\gamma$ coefficient of order
$\gamma\simeq 200 mJ/(mol\cdot K^2)$. The spin susceptibility is Pauli
like and similarly enhanced as $\gamma$, giving rise to a
Sommerfeld-Wilson ratio of order unity. No indication of magnetic
order is found down to $T=0.045 K$, but below $2K$ the susceptibility
starts to increase again indicating the presence of another low energy
scale [126]. These findings strongly suggest heavy-fermion behavior
which is further confirmed by the observation that the ratio
$A/\gamma^{\nu}$ with $\nu=2$ is similar to that of other
heavy-fermion systems. We reemphasize that the $\gamma$ coefficient
exceeds the one, e.g., of $Na$ metal by a factor of more than $10^2$
despite a carrier concentration of only 1 per $10^3\ Yb$ ions. This
shows clearly that the low-energy scale must involve spin degrees of
freedom of the $Yb^{3+}$ ions. Indeed, chains of antiferromagnetically
coupled spins have a linear specific heat $C=\gamma T$. Inelastic
neutron scattering experiments by Kohgi and coworkers [22] have
demonstrated that the magnetic excitation spectrum is that of a
Heisenberg chain with a coupling constant $J=25 K$. This spectrum
leads to a $\gamma$ value of the observed size. The physical origin of
the heavy-fermion excitations is therefore very different here than
that in the Kondo-lattice case. 

A theory has been developed which explains consistently the above
experimental findings [30]. It is based on interpreting the structural
phase transition in terms of a band Jahn-Teller (CBJT)
effect. This interpretation is suggested by the physical
considerations outlined in Sec. IV in connection with the hole
crystallization taking place. In the theory applied here the
crystallization is due to a strong deformation-potential coupling
which is quite common in mixed-valence systems. This potential has its origin
in the Coulomb repulsion of the $4f$ holes. 

The $CBJT$ transition splits the fourfold degenerate quasi-1$d$
density of states into a nondegenerate one corresponding to the short
chains and a threefold one due to the long chains. The nondegenerate
one is lower in energy and would be half filled if charge ordering
were perfect and the holes were uncorrelated fermions. Instead, the
holes are strongly correlated. Two holes on a site imply a $4f^{12}$
configuration for $Yb$ and that has a much too high energy to
occur. Therefore, we are dealing with an almost full lower (hole)
Hubbard band rather than with an almost half-filled conduction
band. Therefore, the system should be close to an insulator. That
$Yb_4As_3$ is a semimetal and not an insulator is most probably
related to the nonvanishing hopping matrix elements between $4f$
orbitals in the long and short chains. We have discussed in Sec. IV
that the zero-point fluctuations of the $4f$ holes lead to a partial
(though small) transfer of holes to the long chains. We speak of
self-doping if this transfer causes the gap in the excitation spectrum
of (quasi) one-dimensional Hubbard chains to vanish. Accurate
conditions for self-doping are not easily worked out, but a first step
in this direction was done recently [127]. 

The phase transition is described by an effective Hamiltonian of the
form of Eq. (4.2).

With increasing charge ordering, correlations become more and more
important because with the increase in concentration of holes in the
short chains their average distance decreases. Therefore, at low
temperatures $T$ the $t-J$ Hamiltonian (see Eq. (3.2)) or a Hubbard
Hamiltonian (3.1) must be used. Using the former and making use of a
slave-boson mean-field approximation we arrive at an effective mass
enhancement of the form 

\begin{equation}
\frac{m^*}{m_b}~=~\frac{t}{t\delta+(3/4)\chi J}.
\end{equation}

Here $m_b$ denotes the band mass,
$\chi=\chi_{ij}=<\sum_{\sigma}f^+_{i1\sigma}f_{j1\sigma}>,\delta$ is
the deviation of the short chains denoted by 1 from half filling and
$J=4t^2/U$, where $U$ is the on-site Coulomb repulsion between
holes. With $U=10 eV$ one finds $J=1\cdot 10^{-3}$ eV and using
$\chi(T=0)=(2/\pi)sin(\pi(1-\delta)/2)$ with $\delta = 10^{-3}$ one
obtains a ratio of $m^*/m_b\simeq 100$. The derivation of the mass
enhancement hides somewhat the fact that spin degrees of freedom are
responsible for the heavy quasiparticles. They become better visible
when one sets $\delta=0$, which is the case of no charge
carriers. Even then one finds fermionic excitations with a large
effective mass and corresponding heat coefficient $\gamma$. The theory
was recently improved [103] by including in the Hamiltonian (5.21) an
on-site Coulomb repulsion $U$ between $4f$ holes. This
one-dimensional Jahn-Teller model can be solved exactly by a Lieb-Wu
Bethe type ansatz. Of particular interest is that a self-doped
distorted phase is obtained in a sizable regime of parameters. Since
spin-wave-like excitations are responsible for the fermionic
low-energy excitations associated with the specific heat and
susceptibility at low $T$ we are dealing here with charge-neutral
heavy fermions in distinction to the charged heavy electrons, which
appear, e.g., in $CeAl_3$. Therefore, we speak of an uncharged or
neutral heavy Fermi liquid. 

The physical interpretation given above allows for an explanation of
another experiment. It has been previously found that an applied
magnetic field of $H=4$ Tesla has little influence on the $\gamma$
coefficient above 2 K, but suppresses $\gamma$ considerably below 2 K
[128]. This effect is unexpected, since one would have thought that
the changes are of order $(\mu_B H/k_B T^*)^2$ and therefore very
small. However, we can explain the experiments by providing for a weak
coupling between parallel short chains. When linear spin-wave theory
is applied, a ratio of order $10^{-4}$ between interchain and
intrachain coupling opens an anisotropy gap which modifies $C(T)$ in
accordance with observation [129].\\ 

\begin{center} {\large\bf Acknowledgement}
\end{center}

\noindent I would like to thank K. Doll, M. Dolg, J. Igarashi,
B. Paulus, B. Schmidt, H. Stoll, P. Thalmeier, P. Unger, V. Zevin and
G. Zwicknagl for years of fruitful discussions and cooperation. 

\newpage

{\Large\bf{References}}

\vspace{1cm}

\begin{enumerate}
\item P. Hohenberg and W. Kohn, Phys. Rev. \underline{\bf 136}, B 864
(1964)

\item W. Kohn and L. J. Sham, Phys. Rev. \underline{\bf 140}, A 1133
(1965)

\item R. O. Jones and O. Gunnarsson, Rev. Mod. Phys. \underline{\bf
61}, 689 (1984)

\item R. M. Dreizler and E. K. U. Gross, {\sl Density Functional
Theory} (Springer, Berlin 1990); see also N. H. March, {\sl Electron
Density Theory of Atoms and Molecules} (Academic Press, London 1992)

\item O. Gunnarsson and B. Lundqvist, Phys. Rev. B \underline{\bf 7},
1912 (199..)

\item D. C. Langreth and J. P. Perdew, Solid State
Comm. \underline{\bf 17}, 1425 (1975)

\item S. F. Boys, Proc. R. Soc. London A \underline{\bf 200}, 542
(1950); see also, e.g., C. E. Dykstra, {\sl Ab initio Calculations of
the Structure and Properties of Molecules} (Elsevier, Amsterdam 1988)

\item J. Cizek, Adv. Chem. Phys. \underline{\bf 14}, 35 (1969)

\item W. Kutzelnigg, in Modern Theoret. Chemistry, Vol. 3, ed. by
H. F. Schaefer III (Plenum, New York 1977)

\item R. Ahlrichs, Comput. Phys. Commun. \underline{\bf 17}, 31 (1979)

\item H. K\"ummel, K. H. L\"uhrmann, J. G. Zabolitzky, Phys. Lett. C
\underline{\bf 36}, 1 (1978)

\item F. Coester and H. K\"ummel, Nucl. Phys. \underline{\bf 17}, 477
(1960)

\item K. Becker and P. Fulde, J. Chem. Phys. \underline{\bf 91}, 4223
(1989)

\item P. Fulde, {\sl Electron Correlations in Molecules and Solids}, 3
rd. edit. (Springer, Heidelberg 1995)

\newpage

\item G. Stollhoff and P. Fulde, J. Chem. Phys. \underline{\bf 73},
4548 (1980) and earlier reference cited  therein

\item P. Pulay, Chem. Phys. Lett. \underline{\bf 100}, 151 (1983); see
also C. Hampel and H.-J. Werner, J. Chem. Phys. \underline{\bf 104},
6286 (1996)

\item A. Lizon-Nordstr\"om and F. Indurain, solid State
Comm. \underline{\bf 94}, 335 (1995)

\item P. O. L\"owdin, J. Mol. Spectrosc. \underline{\bf 10}, 12 (1963)
and \underline{\bf 13}, 326 (1964); see also Int. J. Quantum
Chem. \underline{\bf 21}, 69 (1982)

\item K. Becker and W. Brenig, Z. Phys. B \underline{\bf 79}, 195
(1990)

\item E. Wigner, Phys. Rev. \underline{\bf 46}, 1002 (1934) {\sl and}
Trans. Faraday Soc. \underline{\bf 34}, 678 (1938)

\item A. Ochiai, T. Suzuki and T. Kasuya,
J. Phys. Soc. Jpn. \underline{\bf 59}, 4129 (1990)

\item M. Kohgi, K. Iwasa, A. Ochiai, T. Suzuki. J.-M. Mignon,
B. Gollon, A. Gukasov, J. Schweizer, K. Kakurai, M. Nishi, A. D\"onni
and T. Osakabe (in print)

\item G. R. Stewart, Rev. Mod. Phys. \underline{\bf 56}, 755
(1984)

\item H. R. Ott, Prog. Low Temp. Phys. \underline{\bf 11}, 215
(1987)

\item P. Fulde, J. Keller and G. Zwicknagl in {\sl Solid State
Physics}, Vol. 41, ed. by H. Ehrenreich, D. Turnbull (Academic Press,
San Diego 1988) p. 1

\item N. Grewe and F. Steglich in {\sl Handbook on the Physics and
Chemistry of Rare Earths}, Vol. 14, ed. by K. A. Gschneidner, Jr.,
L. Eyring (North-Holland, Amsterdam 1991)

\item P. Wachter, Handbook on the Physics and Chemistry of Rare
Earths, Vol. 19, ed. by K. A. Gschneidner, Jr., L. Eyring,
G. H. Lander and G. R. Chappin (Elsevier, Amsterdam 1994) p. 177

\item A. C. Hewson, {\sl The Kondo Problem to Heavy Fermions}
(Cambridge University Press, Cambridge 1993)

\newpage

\item P. Fulde, V. Zevin and G. Zwicknagl, Z. Phys. B \underline{\bf
92}, 133 (1993)

\item P. Fulde, B. Schmidt and P. Thalmeier,
Europhys. Lett. \underline{\bf 31}, 323 (1995)

\item K. Andres, J E. Graebner and H. R. Ott,
Phys. Rev. Lett. \underline{\bf 35}, 1779 (1975)

\item T. Brugger, T. Schreiner, G. Roth, P. Adelmann and G. Czjzek,
Phys. Rev. Lett. \underline{\bf 71}, 2481 (1993)

\item T. Suzuki, Phys. Prop. Actinide and Rare Earth
Comp. \underline{\bf 33} AP, Series 8, 267 (1993) 

\item K. Kladko and P. Fulde, (to be published)

\item T. Schork and P. Fulde, J. Chem Phys. \underline{\bf 97}, 9195
(1992)

\item H. Stoll. Phys. Rev. B \underline{\bf 46}, 6700 (1992); {\sl
and} Chem. Phys. Lett. \underline{\bf 191}, 548 (1992)

\item B. Paulus, P. Fulde and H. Stoll, Phys. Rev. B \underline{\bf
51}, 10512 (1995)

\item K. Doll, M. Dolg, P. Fulde and H. Stoll, Phys. Rev. B
\underline{\bf 52}, 4842 (1995)

\item B. Paulus, P. Fulde and H. Stoll, Phys. Rev. B \underline{\bf
54}, 2556 (1996)

\item K. Doll, M. Dolg and H. Stoll, Phys. Rev. \underline{\bf 54},
13529 (1996)

\item MOLPRPO is a package of ab initio programs written by
H.-J. Werner and P. J. Knowles, with contributions from I. Alml\"of,
R. D. Amos, M.-J. O. Deegan, S. T. Elbert, C. Hampel, W. Meyer,
K. Peterson, R. Pitzer, A. J. Stone and P. R. Taylor; the CPP program
was written by A. Nicklass

\item R. Dovesi, C. Pisani and C. Roetti, Int. J. Quantum
Chem. \underline{\bf 17}, 517 (1980); C. Pisani, R. Dovesi and
C. Roethi, Lecture Notes in Chemistry, vol. 48 (Springer, Berlin 1988)

\item J. M. Foster and S. F. Boys, Rev. Mod. Phys. \underline{\bf 32},
300 (1960)

\newpage

\item A. Shukla, M. Dolg, H. Stoll and P. Fulde,
Chem. Phys. Lett. \underline{\bf 262}, 213 (1996)

\item A. Bergner, M. Dolg, W. K\"uchle, H. Stoll and H. Preu\ss\,
Mol. Phys. \underline{\bf 80}, 1431 (1993)

\item G. Igel-Mann, H. Stoll and H. Preu\ss\,
Mol. Phys. \underline{\bf 65}, 1321 (1988)

\item W. A. Harrison, Phys. Rev. B \underline{\bf 23}, 5230 (1981)

\item D. Gl\"otzel, B. Sagall and O. K. Anderson, Solid State
Comm. \underline{\bf 36}, 403 (1980)

\item Y.-M. Juan and E. Kaxiras, Phys. Rev. B \underline{\bf 48},
14944 (1993)

\item Y.-M. Juan, E. Kaxiras and R. G. Gordon, Phys. Rev. B
\underline{\bf 51}, 9521 (1995)

\item S. Fahy, X. W. Wang and S. G. Louie, Phys. Rev. B \underline{\bf
42}, 3503 (1990)

\item W. Borrmann and P. Fulde, Phys. Rev. B \underline{\bf 31}, 7800
(1985)

\item S. Froyens and W. A. Harrison, hys. Rev. B \underline{\bf 20},
2420 (1979)

\item B. Paulus, PHD Thesis, Universit\"at Regensburg (1995)

\item J. Lievin, J. Breulet, P. Clerq and J. Y. Metz,
Theor. Chem. Acta \underline{\bf 61}, 512 (1982)

\item CRC Handbook of Chemistry and Physics, 75th edition, Editor:
David R. Lide (CRC Press, Boca Raton, 1994/95)

\item M. Catti, G. Valerio, R. Dovesi and M. Causa, Phys. Rev. B
\underline{\bf 49}, 14179 (1994)

\item W. C. Mackrodt, N. M. Harrison, V. R. Saunders, N. L. Allan,
M. D. Towler, E. Apra and R. Dovesi, Phil. Mag. A \underline{\bf 68},
653 (1993)

\item M. D. Towler, N. L. Allan, N. M. Harrison, V. R. Saunders,
W. C. Mackrodt and E. Apra, Phys. Rev. B \underline{\bf 50}, 5041
(1994)

\item R. J. Cave, E. R. Davidson, J. Chem. Phys. \underline{\bf 89},
6798 (1988)

\item K. Doll, M. Dolg, P. Fulde and H. Stoll (submitted for
publication)

\item P. W. Atkins, {\it Molecular Quantum Mechanics}, (Oxford Univ. Press,
Oxford 1983)

\item K. Doll (private communication)

\item G. J. M. Janssen and W. C. Nieuwpoort, Phys. Rev. B
\underline{\bf 38}, 3449 (1988); see also C. de Graaf, R. Broer and
W. C. Nieuwpoort, Chem. Phys. \underline{\bf 208}, 35 (1996)

\item M. Takahashi and J. Igarashi, Annalen der Physik, \underline{\bf
5}, 247 (1996); see also F. Manghi, C. Calandra and S. Ossicini,
Phys. Rev. Lett. \underline{\bf 73}, 3124 (1994)

\item see, e. g., V. I. Anisimov, J. Zaanen, O. K. Andersen,
Phys. Rev. B \underline{\bf 44}, 943 (1991)

\item for a review see S. H\"ufner {\sl Photoelectron Spectroscopy},
Springer Ser. Solid-State Sci., Vol. 82 (Springer Verlag, Berlin,
Heidelberg 1995)

\item M. C. Gutzwiller, Phys. Rev. Lett. \underline{\bf 10}, 159
(1963)

\item H. Hubbard, Proc. R. Soc. London A \underline{\bf 276}, 238
(1963)

\item J. Kanamori, Prog. Theor. Phys. \underline{\bf 30}, 275 (1963)

\item A. Brooks Harris and R. V. Lange, Phys. Rev. \underline{\bf
151}, 295 (1967)

\item A. M. Ole\'s and G. Stollhoff, Phys. Rev. B \underline{\bf 29},
314 (1984)

\item M. S. Hybertsen, M. Schl\"uter and N. E. Christensen, Phys. Rev.
B \underline{\bf 39}, 9028 (1989)

\item for a review see, e. g., E. Dagotto,
Rev. Mod. Phys. \underline{\bf 66}, 763 (1994)

\item B. T. Pickup and O. Goscinski, Mol. Phys. \underline{\bf 26},
1013 (1973)

\item J. Linderberg and Y. \"Ohrn, {\sl Propagators in Quantum Chemistry}
(Academic Press, London 1973)

\item S. H\"ufner and G. K. Wertheim, Phys. Lett. \underline{\bf 47A},
349 (1974)

\item D. R. Penn, Phys. Rev. Lett. \underline{\bf 42}, 921 (1979)

\item A. Liebsch, Phys. Rev. Lett. \underline{\bf 43}, 1431 (1979)
{\sl and} Phys. Rev. B \underline{\bf 23}, 5203 (1981)

\item J. Igarashi, J. Phys. Soc. Jpn. \underline{\bf 52}, 2827 (1983);
ibid \underline{\bf 54}, 260 (1985)

\item L. M. Roth, Phys. Rev. \underline{\bf 186}, 1, 428 (1969)

\item J. A. Hertz and D. M. Edwards, J. Phys. \underline{\bf 3}, 2174
(1973); {\sl ibid} \underline{\bf 3}, 2191 (1973)

\item P. Unger, J. Igarashi and P. Fulde, Phys. Rev. B. \underline{\bf
50}, 10485 (1994)

\item D. van der Marel and G. A. Sawatzky,
Phys. Rev. B. \underline{\bf 37}, 10674 (1988)

\item L. D. Faddeev, Zh. Eksp. Teor. Fiz. \underline{\bf 39}, 1459
(1960) [Engl. transl.: Sov. Phys. - JETP \underline{\bf 12}, 1014
(1961)]

\item J. Igarashi, P. Unger, K. Hirai and P. Fulde, Phys. Rev. B
\underline{\bf 49}, 16181 (1994)

\item P. Unger and P. Fulde, Phys. Rev. B \underline{\bf 47}, 8947
(1993); {\sl ibid} B \underline{\bf 48}, 16607 (1993); {\sl ibid} B
\underline{\bf 51}, 9245 (1995)

\item F. C. Zhang and T. M. Rice, Phys. Rev. B \underline{\bf 37},
3754 (1987)

\item W. Stephan and P. Horsch in {\sl Dynamics of Magnetic
Fluctuations in High-Temperature Superconductivity}, ed. by G. Reiter,
P. Horsch, G. Psaltakis (Plenum Press, New York 1990)

\item T. Tohyama and S. Maekawa, Physics C \underline{\bf 191}, 193
(1992)

\item P. Platzmann, Phys. World p. 22, Dec. 1996 

\item J. Durkan, R. J. Elliott and N. H. March,
Rev. Mod. Phys. \underline{\bf 40}, 812 (1968)

\item T. Kasuya, J. Alloys and Compounds, \underline{\bf 192}, 217
(1993) and earlier work cited there

\item C. M. Care and N. H. March, Adv. Phys. \underline{\bf 24}, 101
(1975)

\item E. J. W. Verwey and P. W. Haaymann, Physica \underline{\bf
8}, 979 (1941)

\item N. F. Mott, Phil. Mag. \underline{\bf 6}, 287 (1961)

\item N. F. Mott, {\sl Metal-Insulator Transitions}, (Taylor and
Trancis, London 1990)

\item J. Hubbard, Proc. Roy. Soc. London, A \underline{\bf 281}, 401
(1964)

\item M. Rams, K. Krolas, K. Tomala, A. Ochiai and T. Suzuki,
Hyperfine Interact. \underline{\bf 97/98}, 125 (1996); earlier
Mossbauer studies by B. Bonville, A. Ochiai, T. Suzuki and E. Vincent,
J. Phys. I \underline{\bf 4}, 594 (1994) have shown the existence of
nonequivalent $Yb$ sites.

\item P. Fulde, Annalen der Physik (in print)

\item A. Ochiai, T. Suzuki and T. Kasuya,
J. Magn. Magn. Mater. \underline{\bf 52}, 13 (1985)

\item H. G. von Schnering and V. Grin (private commun.)

\item Y. M. Li, N. d'Ambrumenil and P. Fulde,
Phys. Rev. Lett. \underline{\bf 78}, 3386 (1997)

\item see, e. g., Proceedings of the International Conference on
Strongly Correlated Electron Systems, Physica B \underline{\bf 206 +
207}

\item C. M. Varma and Y. Yafet, Phys. Rev. B \underline{\bf 13}, 2950
(1976)

\item K. Yoshida, Phys. Rev. \underline{\bf 147}, 223 (1966)

\item S. Doniach, Physica B \underline{\bf 91}, 231 (1977)

\item H. Aoki, S. Uji, A. Albessand and Y. Onuki,
Phys. Rev. Lett. \underline{\bf 71}, 2120 (1993)

\item G. Zwicknagl, Adv. Phys. \underline{\bf 41}, 203 (1993)

\item M. N. Norman and D. Koelling, {\sl Handbook on the Physics and
Chemistry of Rare Earths}, Vol. 17, ed. by K. A. Gschneidner Jr.,
L. Eyring, G. H. Lander and G. R. Choppin (Elsevier, Amsterdam 1993)
p. 1

\item G. Zwicknagl, E. Runge and N. E. Christensen, Physica B
\underline{\bf 163}, 97 (1990)

\item P. Fulde and J. Jensen, Phys. Rev. B \underline{\bf 27}, 4085
(1983); see also R. M. White and P. Fulde,
Phys. Rev. Lett. \underline{\bf 47}, 1540 (1981)

\item H. Aoki, S. Uji, A. Albessand and Y. Onuki,
Phys. Rev. Lett. \underline{\bf 71}, 2120 (1993)

\item G. G. Lonzarich, J. Magn. Magn. Mater. \underline{\bf 76 + 77},
1 (1988)

\item C. A. King and G. G. Lonzarich, Physica B \underline{\bf 171},
161 (1991)

\item H. Keiter and J. C. Kimball, Int. J. Magnet. \underline{\bf 1},
233 (1971)

\item H. Kojima, Y. Kuramoto and M. Tachiki, Z. Phys. B \underline{\bf
54}, 293 (1984)

\item N. E. Bickers, Rev. Mod. Phys. \underline{\bf 59}, 845 (1987)

\item N. E. Bickers, D. L. Cox and J. W. Wilkins,
Phys. Rev. Lett. \underline{\bf 54}, 230 (1985)

\item K. Thomala, G. Weschenfelder, G. Czjzek and E. Holland-Moritz, 
J. Magn. Magn. Mater. \underline{\bf 89}, 143 (1990)

\item V. Zevin, G. Zwicknagl and P. Fulde,
Phys. Rev. Lett. \underline{\bf 60}, 2331 (1988)

\item M. Loewenhaupt, A. Metz, N. M. Pyka, D. M. McK Paul, J. Martin,
V. H. M. Dujin, J. J. M. Trause, H. Mutka and W. Schmidt,
Ann. Phys. \underline{\bf 5}, 197 (1996)

\item J. Litterst (private communication)

\item J. Igarashi, K. Murayama and P. Fulde,
Phys. Rev. B. \underline{\bf 52}, 15966 (1995)

\item S. Chakravarty, B. I. Halperin and D. R. Nelson,
Phys. Rev. B. \underline{\bf 39}, 2344 (1989); S. Chakravarty and
R. Orbach, Phys. Rev. Lett. \underline{\bf 64}, 224 (1990)

\item B. Bonville, A. Ochiai, T. Suzuki and J.-M. Mignon, J. Phys. I
\underline{\bf 4}, 594 (1994)

\item S. Blawid, Hoang Anh Tuan and P. Fulde,
Phys. Rev. B. \underline{\bf 54}, 7771 (1996)

\item R. Helfrich, F. Steglich and A. Ochiai (private communication)

\item B. Schmidt, P. Thalmeier and P. Fulde,
Europhys. Lett. \underline{\bf 35}, 109 (1996)

\item R. Pott, G. G\"untherodt, W. Wichelhaus, M. Ohl and H. Bach,
Phys. Rev. B \underline{\bf 27}, 359 (1983)

\item H. H. Davis, I. Bransky and N. M. Tallan, J. Less Comm. Metals
\underline{\bf 22}, 193 (1970)

\item B. Lorenz, phys. stat. sol. (b) \underline{\bf 125}, 375 (1984)

\item D. Ihle and B. Lorenz, phys. stat. sol. (b) \underline{\bf 116},
539 (1983)

\item J. L. Moran-L\'opez and P. Schlottmann,
Phys. Rev. B. \underline{\bf 22}, 1912 (1980)
\end{enumerate}

\newpage

{\Large\bf Figure Captions}

\vspace{1cm}

{\bf Fig. 1:} \ \ \begin{minipage}[t]{13cm}
Detailed presentation of one-body increments for $Ni O$. From
Ref. [61].
\end{minipage}

{\bf Fig. 2:} \ \ \begin{minipage}[t]{13cm}
Two-body oxygen-oxygen increments for $Mg O$. From Ref [38].
\end{minipage}

{\bf Fig. 3:} \ \ \begin{minipage}[t]{13cm}
Van der Waals interaction in ionic crystals. Calculated values within
the CCSD approximation versus values as obtained from London's
equation (2.28). Circles: interactions between positive and negative
ions; crosses: between negative ions. From Ref. [63].
\end{minipage}\\

{\bf Fig. 4:} \ \ \begin{minipage}[t]{13cm}
Contributions of different increments to the binding energy of $Mg
O$. From Ref [63].
\end{minipage}\\

{\bf Fig. 5:} \ \ \begin{minipage}[t]{13cm}
Schematic representation of the $Cu\ 3d_{x^2-y^2}$ and $O\ 2p_{x(y)}$
orbitals which are treated by a 3-band Hubbard Hamiltonian.
\end{minipage}\\

{\bf Fig. 6:} \ \ \begin{minipage}[t]{13cm}
Spectral density of $Ni$ as obtained from a 5-band Hubbard Hamiltonian
(3.4 - 3.6) with $U = 0.56$, $J = 0.22$, $\Delta J = 0.031$ (in units
of the SCF bandwidth) when different approximations are applied:

(a) \ \ \begin{minipage}[t]{12cm}
full spectrum,
\end{minipage}

(b) \ \ \begin{minipage}[t]{12cm}
SCF approximation,
\end{minipage}

(c) \ \ \begin{minipage}[t]{12cm}
correlations included but with $J = \Delta J = 0$,
\end{minipage}

(d) \ \ \begin{minipage}[t]{12cm}
when $\Delta J = 0$,
\end{minipage}

(e) \ \ \begin{minipage}[t]{12cm}
assuming $\Omega = 1$.
\end{minipage}

[From Ref [83]).
\end{minipage}\\

{\bf Fig. 7:} \ \ \begin{minipage}[t]{13cm}
Spectral density of the $Cu-O$ planes:

(a) \ \ \begin{minipage}[t]{12cm}
at half-filling,
\end{minipage}

(b) \ \ \begin{minipage}[t]{12cm}
for 25 \% of hole doping.
\end{minipage}

Oxygen and $Cu$ contributions are shown by dashed and solid
lines, respectively. Parameter values are $U_d = 8$, $U_p = 3$,
$t_{pp} = 0.5$,
$\epsilon_p - \epsilon_d = 4$ in units of $t_{pd}$. Note, that with
hole doping spectral weight is shifted from the upper Hubbard band to
the region close to the Fermi energy (dotted line). (From Ref. [87]).
\end{minipage}\\

{\bf Fig. 8:} \ \ \begin{minipage}[t]{13cm}
Spectral density obtained by diagonalization of a cluster of $(Cu
O_2)_4$:

(a) \ \ \begin{minipage}[t]{12cm}
at half-filling,
\end{minipage}

(b) \ \ \begin{minipage}[t]{12cm}
for 25 \% of hole doping.
\end{minipage}

Parameter values are similar to those in Fig. 7, i.e., $U_d = 8.8$,
$U_p = 4.24$, $t_{pp} = 0.41$, $\epsilon_p - \epsilon_d = 3.37$. Note
the similarities with Fig. 7. $S$ denotes the Zhang-Rice singlet and
$T$ indicates the spin-triplet contribution. Data has been broadened
by a linewidth. From Ref. [90].
\end{minipage}\\

{\bf Fig. 9:} \ \ \begin{minipage}[t]{13cm}

(a) \ \ \begin{minipage}[t]{12cm}
Structure of $Yb_4 As_3$,
\end{minipage}

(b) \ \ \begin{minipage}[t]{12cm}
four families of chains on which the $Yb$ ions are located.
\end{minipage}
\end{minipage}\\

{\bf Fig. 10:} \ \ \begin{minipage}[t]{13cm}
Fermi surfaces for $Ce Ru_2 Si_2$ and $Ce Ru_2 Ge_2$ as derived from
de Haas - van Alphen measurements [113 - 115]. The upper part is due
to
holes and the lower part is due to electrons. In $Ce Ru_2 Ge_2$ the
$4f$ electron of $Ce$ is well localized while in $Ce Ru_2 Si_2$ is
participates in the formation of the Fermi surface. The volume
enclosed by the Fermi surface differs therefore by one electron
resulting in an increase of the hole part and a decrease of the
electron part in $Ce Ru_2 Ge_2$.
\end{minipage}\\

{\bf Fig. 11:} \ \ \begin{minipage}[t]{13cm}
Observation of heavy-fermion excitations in $Nd_{2-x} Ce_x Cu O_4$.

(a) \ \ \begin{minipage}[t]{12cm}
specific heat $C_p (T)$,
\end{minipage}

(b) \ \ \begin{minipage}[t]{12cm}
$C_p (T)/T$,
\end{minipage}

(c) \ \ \begin{minipage}[t]{12cm}
Spin susceptibility for an overdoped sample with $x = 0.2$.
\end{minipage}

From Ref. [32].
\end{minipage}\\

{\bf Fig. 12:} \ \ \begin{minipage}[t]{13cm}
Schematic plot of the quasiparticle bands of $Nd_{2-x} Ce_x Cu O_4$
for $x \neq 0$. The Fermi energy is indicated by a dotted line. Solid
lines: $f$-like excitations, and dashed lines: $d$-like excitations.
\end{minipage}

\newpage

{\Large\bf Table Captions}

\vspace{1cm}

{\bf Table 1:} \ \ \begin{minipage}[t]{13cm}
Cohesive energy per unit cell (in eV) of the elemental semiconductors
in SCF approximation ($E^{coh}_{SCF}$) and with inclusion of
correlations ($E^{coh}_{corr}$). Percentages in parenthesis are with
reference to the experimental values ($E^{coh}_{exp}$) which include
an estimate of the atomic zero-point fluctuations [47]. For
comparison results of other methods are shown: local density
approximation ($E^{coh}_{LDA}$) [48, 49]; generalized gradient
approximation ($E^{coh}_{GGA}$) [50]; quantum Monte Carlo
($E^{coh}_{QMC}$) [51]. From Ref. [37] with corrections included due
to atomic spin-orbit splittings.
\end{minipage}\\

{\bf Table 2:} \ \ \begin{minipage}[t]{13cm}
Cohesive energy (in eV) of the III - V semiconducting compounds. The
notations are the same as in Table 1. From Ref. [37].
\end{minipage}\\

{\bf Table 3:} \ \ \begin{minipage}[t]{13cm}
Lattice constants for the group IV semiconductors in Angstr\"om: in
SCF approximation ($a_{SCF}$); including a core polarization potential
($a_{cpp}$); with inclusion of correlations ($a_{corr}$); Percentages
in parenthesis are with reference to experimental values
($a_{exp}$). One notices a strong influence of core polarization. From
Ref. [37].
\end{minipage}\\

{\bf Table 4:} \ \ \begin{minipage}[t]{13cm}
Bulk modulus of the group IV semiconductors in Mbar in SCF
approximation ($B_{SCF}$); including a core polarization potential
($B_{cpp}$); with correlations included ($B_{corr}$); Percentages in
parenthesis are with reference to experimental values
($B_{exp}$). From Ref. [37].
\end{minipage}\\

{\bf Table 5:} \ \ \begin{minipage}[t]{13cm}
Parameters in the BOA as obtained by fitting the results for a single
and for neighboring bonds to those of the ab initio calculation with a
minimal basis set. $\delta E^{inter}_{corr}$ denotes those
contributions which come from more distant than nearest-neighbor
bonds. From Ref. [54].
\end{minipage}\\


{\bf Table 6:} \ \ \begin{minipage}[t]{13cm}
Cohesive energy (in eV) of three oxides. The notation is the same as
in Table 1. $E^{coh}_{exp}$ from Ref. [56], $E^{coh}_{SCF}$ from
Ref. [57] ($Mg O$), [58] ($Ca O$) and [59] ($Ni O$). Calculations for
$E^{coh}_{corr}$ are on a CCSD(T) level ($Mg O, Ca O$) and by applying
quasidegenerate variational perturbation theory (QDVPT) [60].
\end{minipage}\\

{\bf Table 7:} \ \ \begin{minipage}[t]{13cm}
Lattice constants (in $\AA$) of three oxides. Same notation as in
Table 3. For $a_{SCF}$ see Refs. [57 - 59].
\end{minipage}\\

{\bf Table 8:} \ \ \begin{minipage}[t]{13cm}
One-body increments for $Mg O, Ca O$ and $Ni O$ in eV. From Refs. [38,
40, 61].
\end{minipage}\\

{\bf Table 9:} \ \ \begin{minipage}[t]{13cm}
Sums of different two-body increments in eV. The last line gives the
sum of all 2-body increments. From Refs. [38, 40, 61].
\end{minipage}\\

{\bf Table 10:} \ \ \begin{minipage}[t]{13cm}
Measured [114] and calculated [111] mass ratios for $Ce Ru_2
Si_2$. Shown are extremal areas of the Fermi surface (in megagauss)
and the effective mass ratios $m^*/m_0$. Unlike the LDA, the
renormalized band theory (RB) reproduces the large measured mass
anisotropies.
\end{minipage}

\newpage

\begin{table}[hbt]
\begin{center}
\begin{tabular}{|c||c|c|c|c|}
\hline
&C&Si&Ge&$\alpha$-Sn\\
\hline
$E_{\rm SCF}^{\rm coh}$&10.74&6.18&4.25&3.65\\
&(71\%)&(66\%)&(53\%)&(53\%)\\
\hline
$E_{\rm corr}^{\rm coh}$&14.36&8.84&7.02&6.13\\
&(95\%)&(94\%)&(88\%)&(90\%)\\
\hline
$E_{\rm exp}^{\rm coh}$&15.10&9.39&8.00&6.83\\
\hline
$E_{\rm LDA}^{\rm coh}$&17.25&10.59&9.06&---\\
\hline
$E_{\rm GGA}^{\rm coh}$&---&8.79&6.83&---\\
\hline
$E_{\rm QMC}^{\rm coh}$&14.90&9.76&---&---\\
\hline
\end{tabular}
\end{center}
Table 1
\end{table}

\begin{table}[hbt]
\begin{center}
\begin{tabular}{|c||c|c|c|c|c|c|}
\hline
&BN&BP&BAs&AlP&AlAs&AlSb\\
\hline\hline
$E_{\rm SCF}^{\rm coh}$&9.09&6.26&5.50&5.39&4.71&3.97\\
&(67\%)&(60\%)&&(64\%)&(59\%)&(60\%)\\
\hline
$E_{\rm corr}^{\rm coh}$&12.38&9.36&8.57&7.95&7.18&6.31\\
&(91\%)&(90\%)&&(95\%)&(90\%)&(95\%)\\
\hline
$E_{\rm exp}^{\rm coh}$&13.61&10.39&---&8.41
&8.00&6.61\\
\hline
\hline
&GaP&GaAs&GaSb&InP&InAs&InSb\\
\hline\hline
$E_{\rm SCF}^{\rm coh}$&4.00&3.54&2.97&3.86&3.51&3.18\\
&(53\%)&(53\%)&(49\%)&(57\%)&(54\%)&(55\%)\\
\hline
$E_{\rm corr}^{\rm coh}$&6.69&6.20&5.39&6.37&5.96&5.47\\
&(91\%)&(93\%)&(89\%)&(94\%)&(92\%)&(94\%)\\
\hline
$E_{\rm exp}^{\rm coh}$&7.37&6.69&6.07&6.80&6.50&5.80\\
\hline
\end{tabular}
\end{center}
Table 2
\end{table}

\begin{table}[hbt]
\begin{center}
\begin{tabular}{|c||c|c|c|c|}
\hline
&C&Si&Ge&Sn\\
\hline\hline
$a_{\rm SCF}$&3.5590&5.4993&5.7516&6.6001\\
&(-0.2\%)& (+1.2\%)& (+1.7\%)& (+1.7\%)\\
\hline
$a_{\rm cpp}$&---&5.4662&5.6653&6.4549\\
&& (+0.6\%)& (+0.2\%)& (-0.5\%)\\
\hline
$a_{\rm corr}$&3.5833&5.4256&5.6413&6.4443\\
&(+0.5\%)& (-0.1\%)& (-0.3\%)& (-0.7\%)\\
\hline
$a_{\rm exp}$&3.5657&5.4317&5.6575&6.4892\\
\hline
\end{tabular}
\end{center}
Table 3
\end{table}

\begin{table}[hbt]
\begin{center}
\begin{tabular}{|c||c|c|c|c|}
\hline
&C&Si&Ge&Sn\\
\hline\hline
$B_{\rm SCF}$&4.815&1.038&0.961&0.638\\
&(+9\%)&(+5\%)&(+31\%)&(+20\%)\\
\hline
$B_{\rm cpp}$&---&1.009&0.889&0.562\\
&&(+2\%)&(+21\%)&(+6\%)\\
\hline
$B_{\rm corr}$&4.196&0.979&0.711&0.510\\
&(-5\%)&(-1\%)&(-3\%)&(-4\%)\\
\hline
$B_{\rm exp}$&4.42&0.99&0.734&0.531\\
\hline
\end{tabular}
\end{center}
Table 4\\
\end{table}

\begin{table}[hbt]
\begin{center}
\begin{tabular}{|c||c|c|c|c|}
\hline
&C&Si&Ge&Sn\\
\hline\hline
bandwidth&46.0&17.8&17.2&13.8\\[-2ex]
$[53]$&&&&\\
\hline
$t_0$&8.34&4.14&4.01&3.20\\
$V_0$&3.41&1.78&1.44&1.25\\
$V_1$&0.87&0.51&0.45&0.38\\
\hline
$E_{\rm corr}^{\rm inter}$ (BOA)&-2.06&-1.19&-0.91&-0.83\\
\hline
$\delta E_{\rm corr}^{\rm inter \ }$
 &-0.45&-0.42&-0.40&-0.38\\
\hline
\end{tabular}
\end{center}
Table 5
\end{table}

\begin{table}[hbt]
\begin{center}
\begin{tabular}{|c||c|c|c|}
\hline
&MgO&CaO&NiO\\
\hline\hline
$E_{SCF}^{coh}$&5.85&5.92&4.37\\
\hline
$E_{corr}^{coh}$&7.82&8.06&7.00\\
&(96 \%)&(93 \%)&(93 \%)\\
\hline
$E_{exp}^{coh}$&8.14&8.65&7.49\\
\hline
\end{tabular}
\end{center}
Table 6
\end{table}

\begin{table}[hbt]
\begin{center}
\begin{tabular}{|c||c|c|c|}
\hline
&MgO&CaO&NiO\\
\hline\hline
$a_{SCF}$&4.191&4.864&4.264\\
\hline
$a_{corr}$&4.184&4.801&4.164\\
\hline
$a_{exp}$&4.207&4.803&4.170\\
\hline
\end{tabular}
\end{center}
Table 7
\end{table}

\begin{table}[hbt]
\begin{center}
\begin{tabular}{|c||c|c|c|}
\hline
&MgO&CaO&NiO\\
\hline\hline
lattice const.&4.21&4.81&4.17\\
\hline
$O \rightarrow O^{2-}$&-2.04&-2.05&-2.14\\
\hline
$X \rightarrow X^{2+}$&0.99&1.00&1.58\\
\hline
sum of one-body&-1.05&-1.05&-0.56\\[-2ex]
increments&&&\\
\hline
\end{tabular}
\end{center}
Table 8
\end{table}

\begin{table}[hbt]
\begin{center}
\begin{tabular}{|c||c|c|c|}
\hline
&MgO&CaO&NiO\\
\hline\hline
$O - O$&-0.36&-0.15&-0.56\\
\hline
$X - O$&-0.41&-0.78&-1.56\\
\hline
$X - X$&-0.002&-0.02&-0.04\\
\hline\hline
sum of two-body&-0.77&-0.96&-2.15\\[-2ex]
increments&&&\\
\hline
\end{tabular}
\end{center}
Table 9
\end{table}

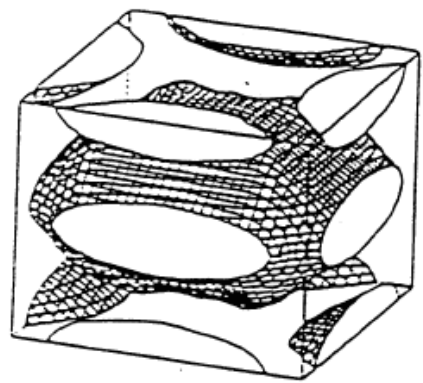
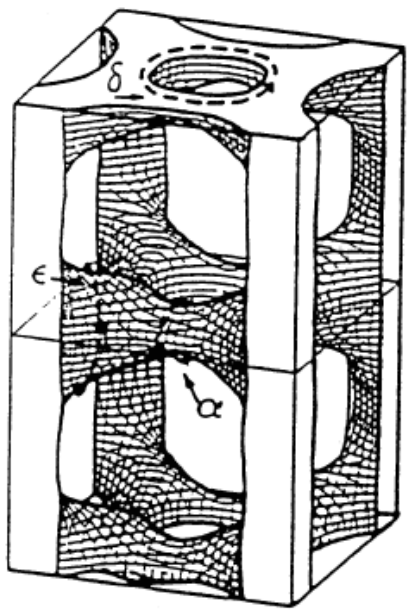
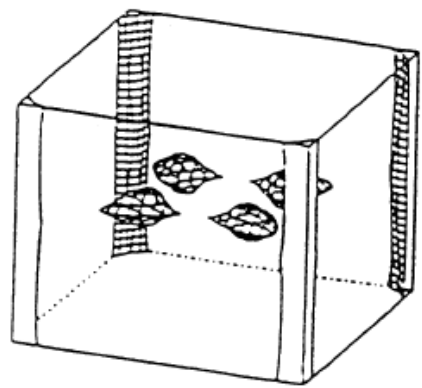
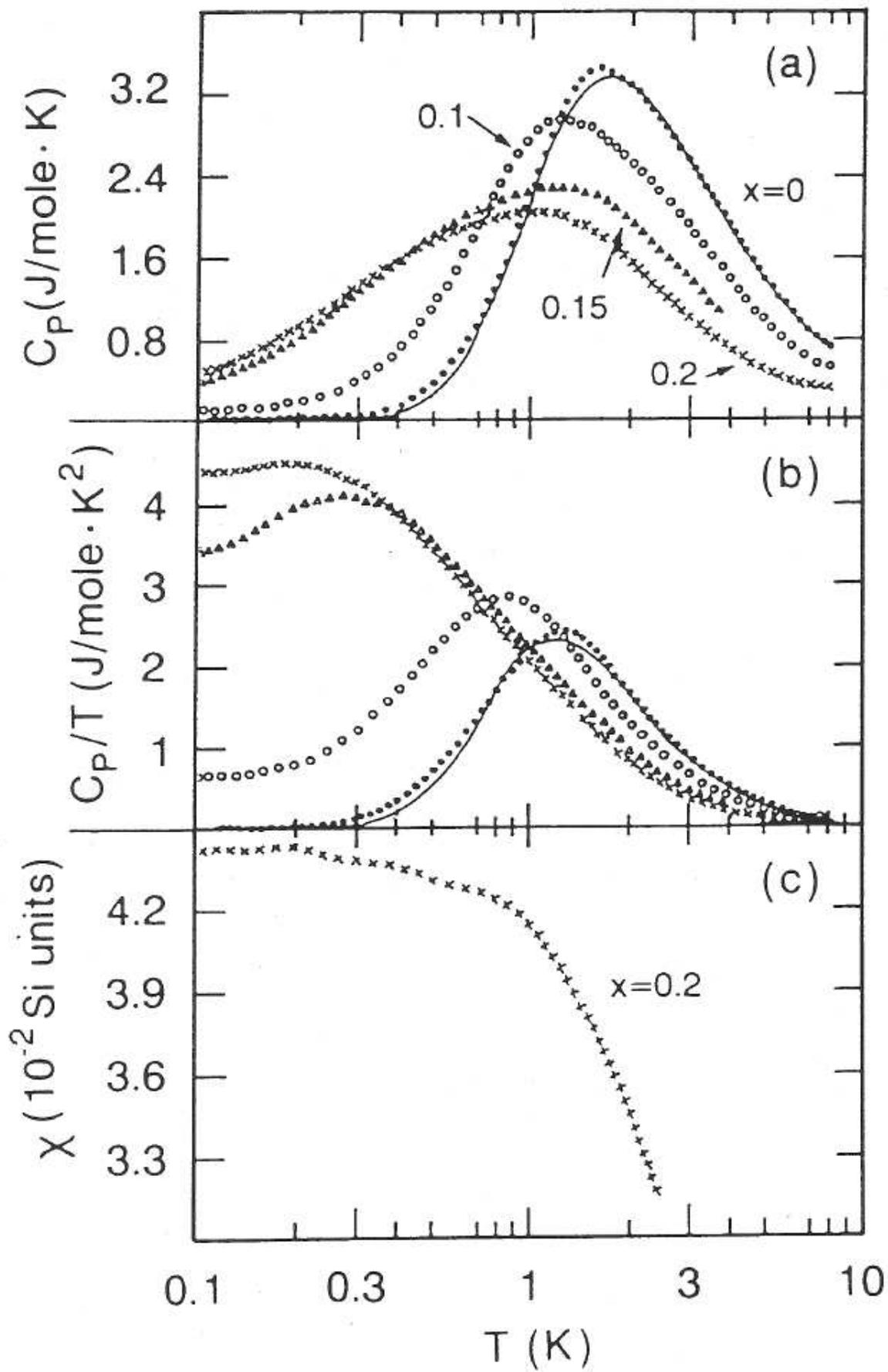
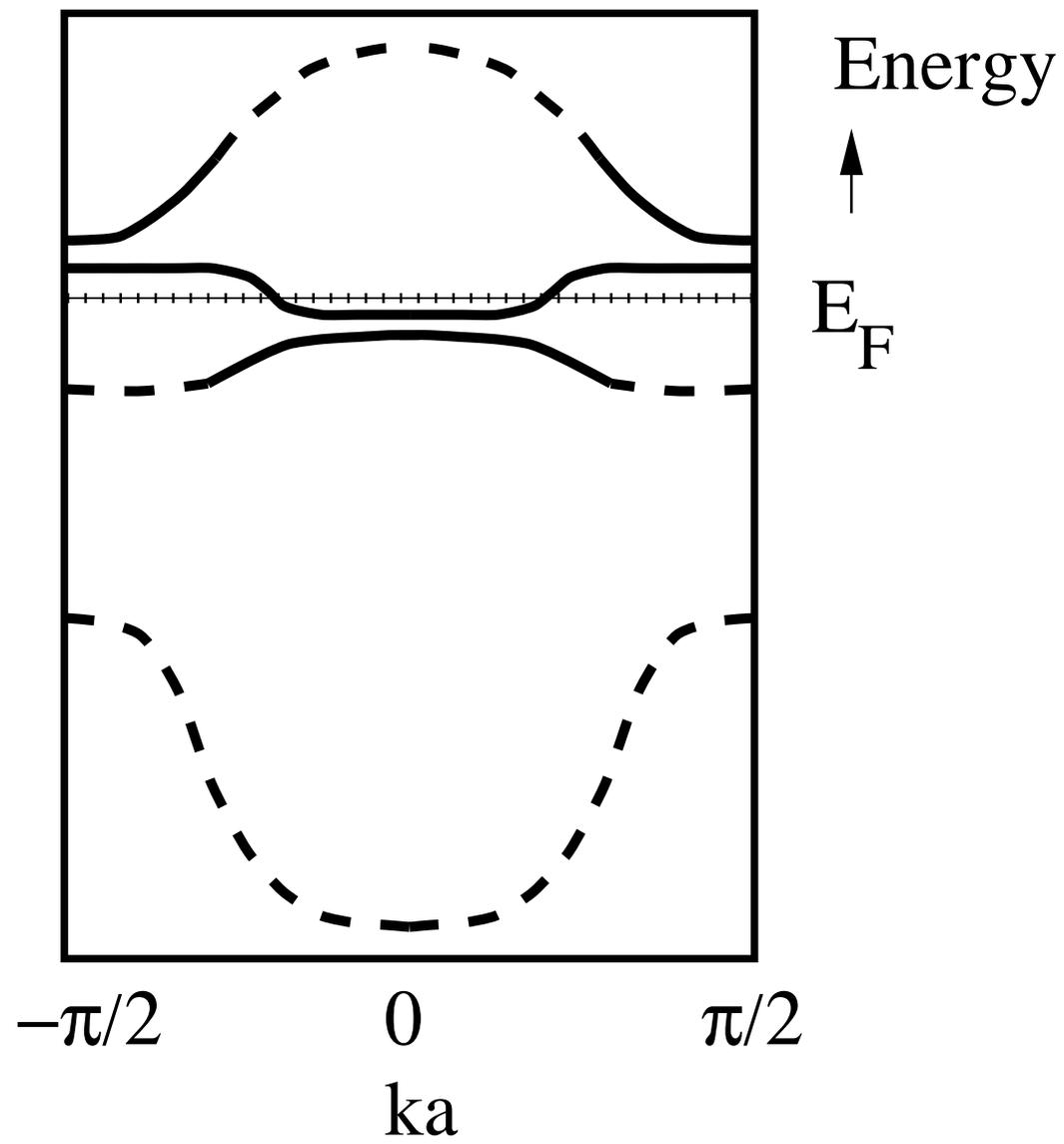
\end{document}